\documentclass{aa}  

\usepackage{graphicx}
\usepackage{subfigure}
\usepackage{xcolor}
\usepackage{soul}
\usepackage{txfonts}

\newcommand\mancha{\textsc{Mancha3D~}}

\usepackage{hyperref}
\begin{document}

   \title{Observational and numerical characterization of a recurrent arc-shaped front propagating along a coronal fan}
   \titlerunning{Propagating front along a coronal fan}
   \author{M.V. Sieyra\inst{1} \and S. Krishna Prasad\inst{1} \and G. Stenborg\inst{2} \and E. Khomenko\inst{3,4} \and T. Van Doorsselaere\inst{1} \and A. Costa\inst{5} \and A. Esquivel\inst{6} \and J.M. Riedl\thanks{Formerly at Centre for mathematical Plasma Astrophysics, Department of Mathematics, KU Leuven}
          }

   \institute{Centre for mathematical Plasma Astrophysics, Department of Mathematics, KU Leuven, Celestijnenlaan 200B bus 2400, B-3001 Leuven, Belgium\\ 
   \email{vale.sieyra@kuleuven.be}
   \and Johns Hopkins University Applied Physics Laboratory, Laurel, MD, USA\\
   \email{guillermo.stenborg@jhuapl.edu}
   \and Instituto de Astrof\'{\i}sica de Canarias, E-38205 La Laguna, Tenerife, Spain;
   \and  Departamento de Astrof\'{\i}sica, Universidad de La Laguna, E-38205, La Laguna, Tenerife, Spain
   \and Instituto de Astronom\'{\i}a Te\'orica y Experimental, CONICET-UNC, C\'ordoba, Argentina.
   \and Instituto de Ciencias Nucleares, Universidad Nacional Aut\'onoma de M\'exico, A. Postal. 70-543, 04510, Ciudad de M\'exico, M\'exico.}

   \date{Received; accepted}

 
  \abstract
  {Recurrent, arc-shaped intensity disturbances were detected by extreme-ultraviolet channels in an active region. The fronts were observed to propagate along a coronal loop bundle rooted in a small area within a sunspot umbra. Previous works have linked these intensity disturbances to slow magnetoacoustic waves that propagate from the lower atmosphere to the corona along the magnetic field.}
  {The slow magnetoacoustic waves propagate at the local cusp speed, equivalent to the sound speed in a low $\beta$ regime plasma. However, the measured propagation speeds from the intensity images are usually smaller as they are subject to projection effects due to the inclination of the magnetic field with respect to the line-of-sight. Here, we aim to understand the effect of projection by comparing observed speeds with those from a numerical model.}
  {Using multi-wavelength data we determine the periods present in the observations at different heights of the solar atmosphere through Fourier analysis. We calculate the plane-of-sky speeds along one of the loops from the cross-correlation time lags obtained as a function of distance along the loop. We perform a 2D ideal magnetohydrodynamic simulation of an active region embedded in a stratified atmosphere. We drive slow waves from the photosphere with a 3 minutes periodicity. Synthetic time-distance maps are generated from the forward-modelled intensities in coronal wavelengths and the projected propagation speeds are calculated.}
  {The intensity disturbances show a dominant period between [2--3] minutes at different heights of the atmosphere. The apparent propagation speeds calculated for coronal channels exhibit an accelerated pattern with values increasing from 40 to 120 km\,s$^{-1}$ as the distance along the loop rises. The propagation speeds obtained from the synthetic time-distance maps also exhibit accelerated profiles within a similar range of speeds.}
   {We conclude that the accelerated propagation in our observations is due to the projection effect.}


   \maketitle
%

\section{Introduction}
\label{s:intro}
The study of MHD waves has two major applications: coronal seismology and coronal heating. Regarding the first aspect, the properties of MHD waves depend on the properties of the plasma environment where they propagate, therefore it is possible to use waves for probing macroscopic parameters of the plasma in the vicinity of the wave propagation region \citep[e.g.][]{demoortel2012}. In relation to the second point, waves could carry the energy through different layers of the solar atmosphere and deposit it in the corona via different dissipative processes  \citep[see][]{vandoorsselaere2020}, contributing to the maintenance of high temperatures in the corona. 

In this study we focus on propagating slow magnetoacoustic waves. These compressive waves are in general related to periodic intensity perturbations mostly observed in extreme-ultraviolet (EUV) or X-ray emission moving along the field aligned plasma. They have been detected in several coronal structures such as polar plumes \citep[e.g.][]{ofman1997,deforest1998} and active region fan loops \citep[e.g.][]{berghmans1999,demoortel2002,marsh2006}. However there is an alternative interpretation of the observed intensity disturbances based on the data from the EUV Imaging Spectrometer (EIS) onboard Hinode and the Interface Region Imaging Spectrograph (IRIS) that associate them  with intermittent outflows and spicules produced at loop footpoint regions \citep[for more details see][]{depontieu2010,wang2016}. Numerical efforts have been made to tackle this issue in, e.g., \citet[][]{demoortel2015}. There, the authors performed forward modeling of the coronal emission from simulations using wave and periodic outflow drivers, and did not find any significant difference in the synthetic observational characteristics between the different drivers. In the current article, we deal with propagating intensity perturbations observed in active region fan loops, especially the 3 minute oscillations. Such perturbations are considered to be relatively less ambiguous in their interpretation because of their regular periodic pattern lasting for hours to days, and because of their possible connection with the upward propagating 3 minute waves in the lower atmosphere. Therefore, we assume that they are the observational signatures of slow waves and hence we revisit some of the properties related to them. 

Slow magnetoacoustic waves exhibit periods ranging from 2 to 10 minutes. In particular, in loops rooted in a sunspot umbra, their periodicity is about 3 minutes, and, for those rooted in the plage regions it is about 5 minutes \citep[][]{demoortel2002}. There is also a distinction on the periodicity found in active regions and in polar regions \citep[for a detailed discussion see][and references therein]{banerjee2021}. The  amplitude of the perturbations corresponds to a few percent of the background and their phase speeds range from 40 to 200 km\,s$^{-1}$, values that are comparable to the local sound speed. They also decay rapidly with height, with typical decay length scales of the order of tens of megameters \citep[][]{nakariakov2020}.

Slow waves are thought to originate in the lower atmosphere \citep{sych2009,botha2011,jess2012,krishna2015}. There are different hypotheses on this, especially for the 3 minutes waves, such as: externally driven by the global (broad-band) photospheric \textit{p}-modes \citep[][]{marsh2006,jess2012,krishna2015,zhao2016}, magnetoconvection \citep[e.g.][]{jess2012a,chae2017,cho2019} and excitation above the cut-off frequency \citep[][]{fleck1991,bogdan2000}. The generation and propagation of slow waves has also been the subject of many numerical studies. In particular, the mode conversion and energy transport have received considerable attention in 2D \citep{khomenko2009,fedun2011,santamaria2015} and 3D \citep{felipe2010,mumford2015,riedl2021} simulations. In these works they obtained that slow waves, that propagate along the magnetic field, reach the upper atmosphere and the fast waves are reflected back, resulting in more acoustic energy propagating into the corona. Nevertheless, according to \citet{riedl2021}, only about 2\% of the initial energy from the driver reaches the corona because of the strong damping due to the cut-off and to the geometric effects.

All these studies bring us to an important question about the slow magnetoacoustic waves, i.e., what is the role of the magnetic field in helping them reach coronal heights?
The magnetic field plays an important role on the propagation of slow waves, not only acting as a waveguide, but also modifying the cut-off frquency \citep[][]{zhukov2002}. There is a strong consensus that the magnetic field inclination plays a crucial role in the frequency determination of, e.g., penumbral waves because of the consequent reduction of the cut-off frequency \citep{reznikova2012,reznikovaandshibasaki2012,kobanov2013,jess2013,yuan2014}. Recently \citet{zurbriggen2020} showed, through an analytical calculation, that the cut-off frequency of slow magnetoacoustic-gravity waves not only depends on the inclination of the magnetic field, but also of its intensity. The analytical results showed a good agreement with the dominant periods found in an active region. Therefore, the cut-off frequency can be used as a seismological tool to determine the magnetic field and the temperature stratification. 

Another seismological tool is the phase speed of the propagating disturbances in the plane-of-sky (POS). A typical method for studying them is via the so-called time-distance plot.  Applying this technique to the stereoscopic  observations of a coronal loop from the Extreme-Ultraviolet Imager instruments (EUVI) from STEREO A and B spacecraft, \citet{marsh2009} estimated the true propagation phase speed from each spacecraft. The obtained values were $132.0^{+9.9}_{-8.5}~\mathrm{km\,s^{-1}}$ and $132.2^{+13.8}_{-8.6}~\mathrm{km\,s^{-1}}$. Assuming that these waves correspond to slow modes and under the condition of low plasma $\beta$, the corresponding plasma temperatures were estimated as $0.84^{+0.13}_{-0.11}~\mathrm{MK}$ and $0.84^{+0.18}_{-0.11}~\mathrm{MK}$, respectively. Shortly after, \citet{marsh2009b} used spectroscopic diagnostic methods with data from Hinode/EIS and confirmed the temperature that was determined seismologically in the previous work, strengthening the slow mode interpretation. Almost at the same time \citet{wang2009} determined the projected and Doppler speeds of propagating disturbances along a fan-like coronal structure observed in the 195~\AA{} line using the same instrument. With these measurements and applying linear wave theory they obtained the inclination of the magnetic field as $59\pm 8^{\circ}$, a true propagation speed of $128 \pm 25~\mathrm{km\,s^{-1}}$ and the corresponding temperature of the loop as $0.7 \pm 0.3~\mathrm{MK}$ near the footpoint. \cite{yuan2012} designed several techniques to measure the apparent propagation speeds from the time-distance maps in coronal imaging in a systematic way. They applied the techniques to quasi-periodic EUV disturbances propagating at a coronal fan-structure of an active region observed in the 171~\AA{} bandpass from Atmospheric Imager Assembly (AIA) onboard SDO, resulting in values that range from 48 to 52~km\,s$^{-1}$.

In all the above mentioned works the observed apparent speed was constant. On the other hand, \citet{sheeley2014}, employing a running difference technique to track proper motions in XUV disturbances, detected accelerated patterns, from $[35-45]~\mathrm{km\,s^{-1}}$ to $[60-100]~\mathrm{km\,s^{-1}}$, in 2.6 minutes sunspot waves at coronal heights. They attributed this accelerated behaviour to the fading length of the height-time track as well as variations in the inclination of the plume above the sky plane. Also \cite{krishna2017} found accelerating slow magnetoacoustic waves along a coronal loop that show differential propagation speeds in two distinct temperature channels, namely, the 131~\AA{} and 171~\AA{} from SDO/AIA. Using the inclination of the loop with respect to the LOS (obtained from nonlinear force-free magnetic field extrapolations) they deprojected the measured phase speeds and estimated the corresponding local plasma temperatures. They found that the deprojected speeds still show an accelerated profile and they are different for the two channels. Furthermore, the temperature in both  channels increases with the distance along the loop and they found an appreciable difference between the two channels. They suggested that these results imply a multi-thermal, and consequently multi-stranded, structure of the loop. 

Having these findings in mind, in this work we analyse an event consisting of semi-circular, recurrent, arc-shaped intensity fronts that popped out of NOAA AR 11243 on 2011 July 6. The intensity disturbances were captured by SDO/AIA in several EUV channels. It was observed to persist for at least the first 8 hours of the day, with neither flare activity nor coronal mass ejections reported nearby prior or during the occurrence of the phenomenon. A periodicity analysis reveals the presence of a 2.5 minutes period at different heights of the AR near the footpoint of the loop rooted in the sunspot umbra. Kinematic analysis performed on EUV channels of SDO/AIA indicate an accelerated profile along the loop. This event was first analyzed by \citet{stekel2014}, suggesting that the observed acceleration could be due to projection effects and the coronal oscillations may be originated in an umbral dot. With the aim of studying the projection effect caused by the inclination of the magnetic field, we perform numerical simulations. We model the coronal oscillations in a sunspot magnetic field configuration embedded in a gravitationally stratified atmosphere by inducing perturbations with an oscillatory driver near the photosphere. We calculate the projected propagation speeds from the forward-modelled intensity emission and compare the results with those from observations. 

The paper is organized as follows. In Sect.~\ref{s:obs}, we describe the observations and give a detailed account of the observed event along with the periodicity analysis (Sect.~\ref{ss:periods:s:obs}) and the characterization of wave propagation (Sect.~\ref{ss:kin:s:obs}). In Sect.~\ref{s:num}, we introduce the numerical model and in Sect.~\ref{s:results} we display the results obtained from the synthetic intensity emission. We present a discussion on both the observational and numerical results in Sect.~\ref{s:discussion}. Finally, we summarize our results and conclusions in Sect.~\ref{s:conclusions}.
\section{Observations}
\label{s:obs}
For this study we consider the space-based data from the Atmospheric Imaging Assembly \citep[AIA,][]{aia} and the Helioseismic and Magnetic Imager \citep[HMI,][]{hmi} both onboard the Solar Dynamics Observatory \citep[SDO,][]{sdo}. The AIA observations consist of cropped images of  $124\times80$ arcsec$^2$ in five EUV bands (304~\AA{}, 131~\AA{}, 171~\AA{}, 193~\AA{} and 211~\AA{}) and two UV bands (1700~\AA{} and 1600~\AA{}) from NOAA AR 11243 on 2011 July 6. The time sequence starts from 00:00 UT until 02:00 UT with a temporal cadence of 12 s for the images taken in the EUV channels and 24 s for those in the UV spectrum. The leading spot of the active region is located at 14$^\circ$~N and 34$^\circ$~W at the beginning of the day ($\mu\approx 0.83$). The plate scale is 0.6 arcsec for both EUV and UV images. The HMI data correspond to the continuum intensity (near Fe I 6173~\AA{}) and the LOS magnetogram, with a cadence of 45 s and a plate scale of 0.5 arcsec. All the images have been co-aligned, corrected for differential rotation and brought to a common plate scale using the standard AIA software package available in IDL Solarsoft. 

\subsection{Data Analysis}
\label{ss:analysis:s:obs}
\begin{figure*}[t]
\centering
\includegraphics[width=\textwidth]{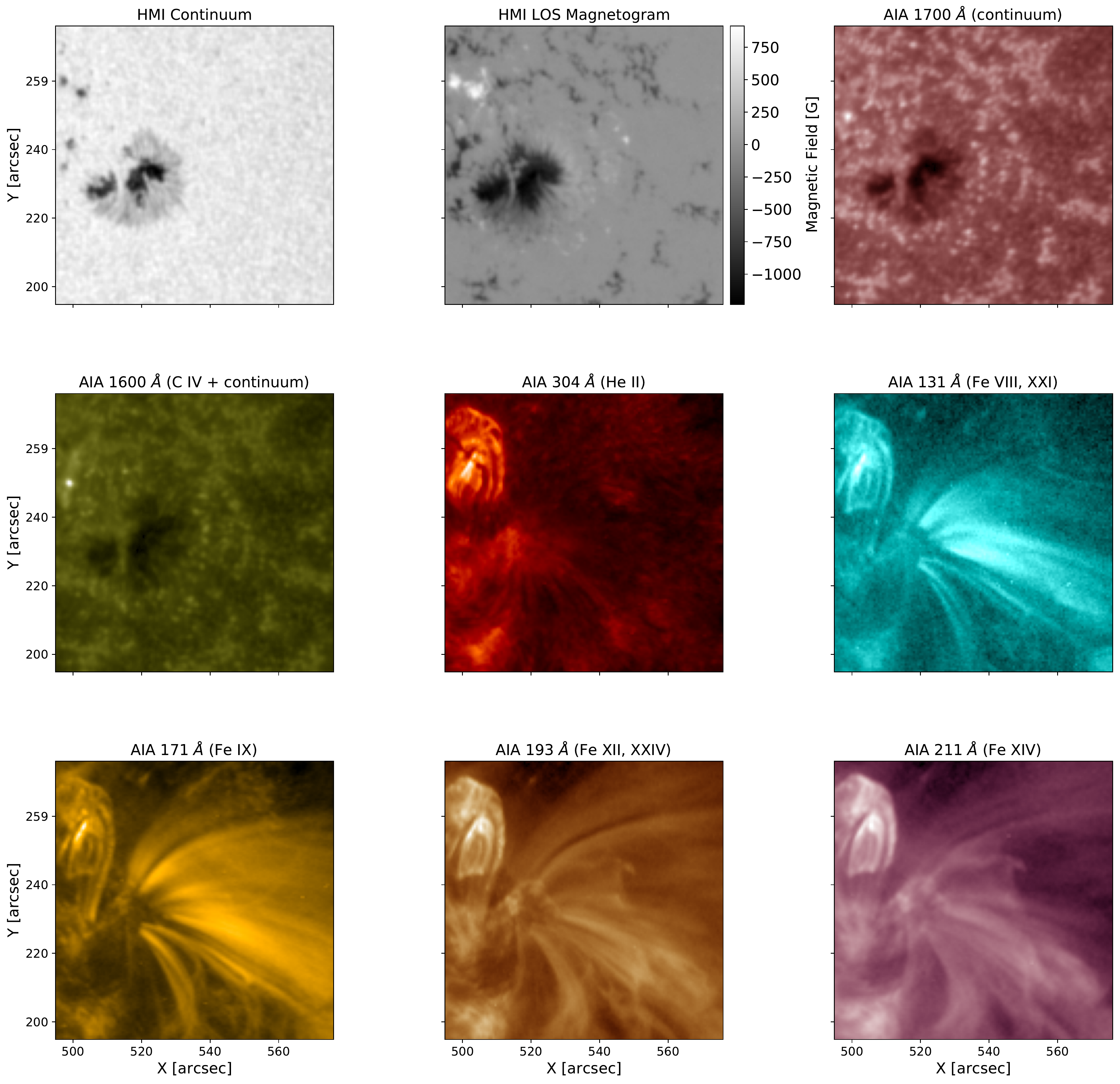}
\caption{Images of NOAA AR 11243 observed in HMI Continuum, HMI LOS Magnetogram and AIA channels. The HMI and AIA UV channels show the sunspot associated with the active region while the coronal EUV channels exhibit the fan-like loop structure extending towards the right side. The primary ions of individual channels are listed in the title of each panel following \citet{aia}}. 
\label{fig:roi}
\end{figure*}
While inspecting a 
time-lapse sequence of 171~\AA{} AIA images taken during the first 8 hours of 2011 July 6, we found a train of semi-circular and recurrent intensity fronts apparently popping out from NOAA AR 11243 towards west (see movie in the on-line version of the journal). The arc-shaped, recurrent fronts appear to propagate through a bundle of loops apparently rooted in the umbra of the sunspot. Figure~\ref{fig:roi} shows snapshots of the active region as recorded by HMI and AIA on 2011 July 6, just at the beginning of the day. 
In the HMI images (top row, left and middle panels) and in the UV images of AIA (1700~\AA{}, top-right panel, and 1600~\AA{}, middle-left panel) of Fig.~\ref{fig:roi} we can see the sunspot associated to NOAA AR 11243. The HMI LOS Magnetogram shows the value of the magnetic field intensity. The coronal channels (131~\AA{}, in the right-middle panel and the 171~\AA{}, 193~\AA{} and 211~\AA{} in the bottom row) show the fan-like loop structure extending from the centre to the right of each snapshot. No flare activity or coronal mass ejections were reported in association with this phenomenon. 

\subsection{Periodicity Analysis} 
\label{ss:periods:s:obs}
\begin{figure*}
    \centering
    \includegraphics[width=0.85\textwidth]{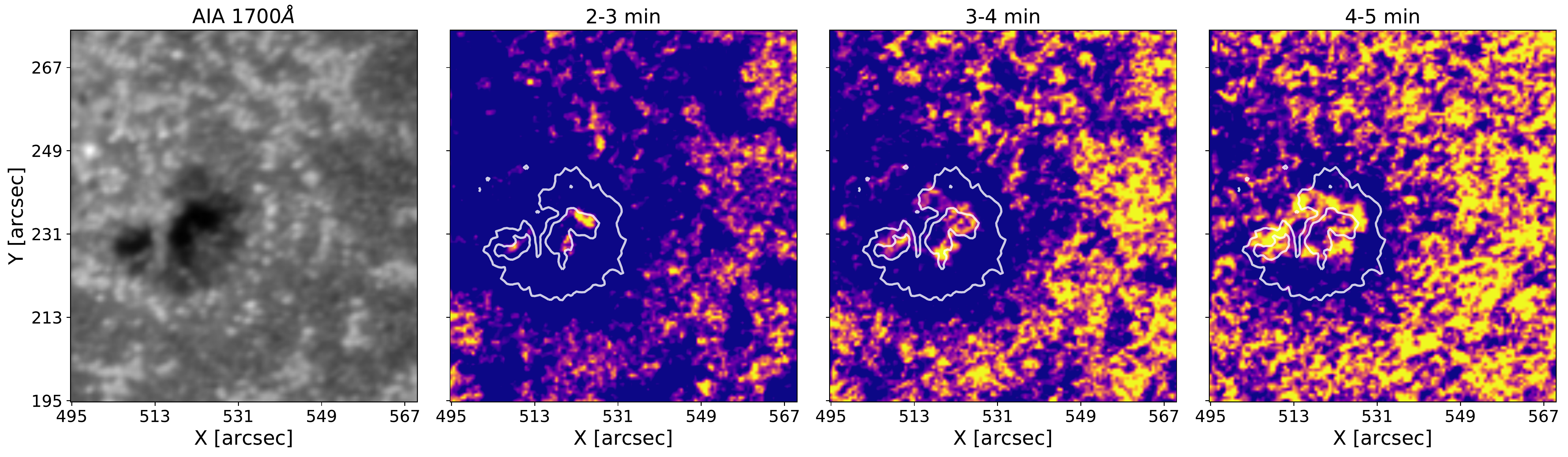}
    \includegraphics[width=0.85\textwidth]{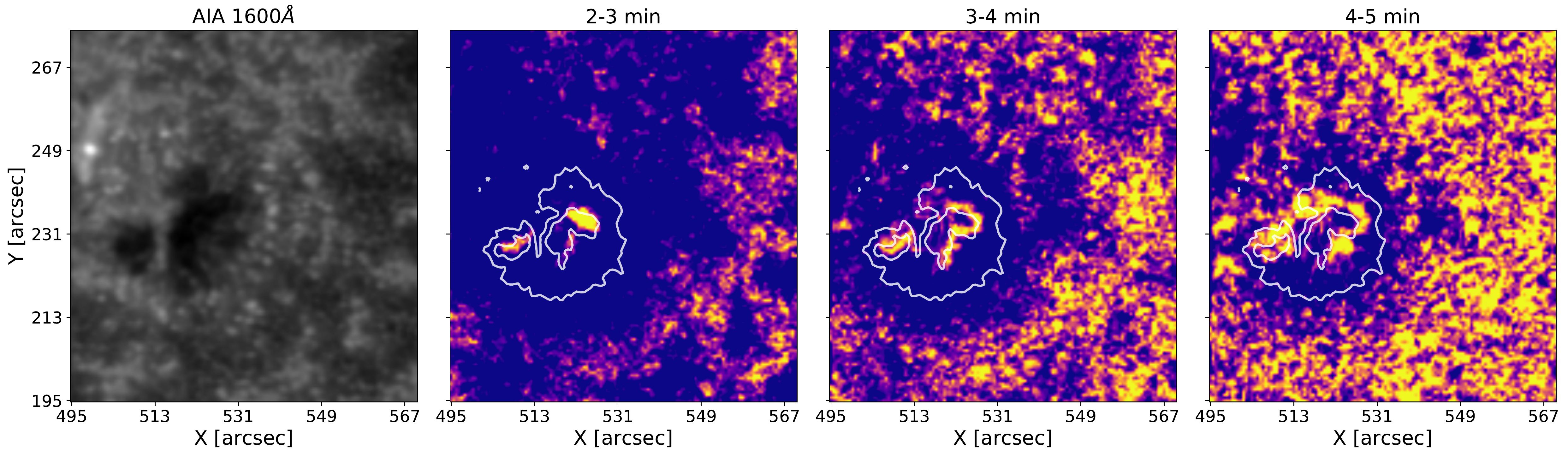}
    \includegraphics[width=0.85\textwidth]{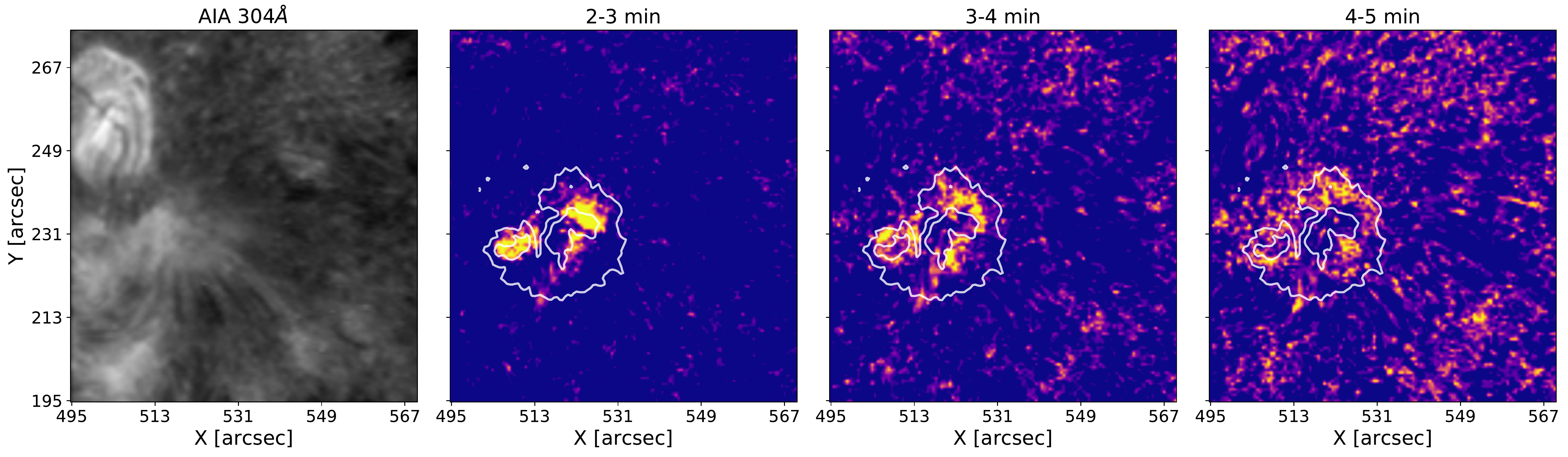}
    \includegraphics[width=0.85\textwidth]{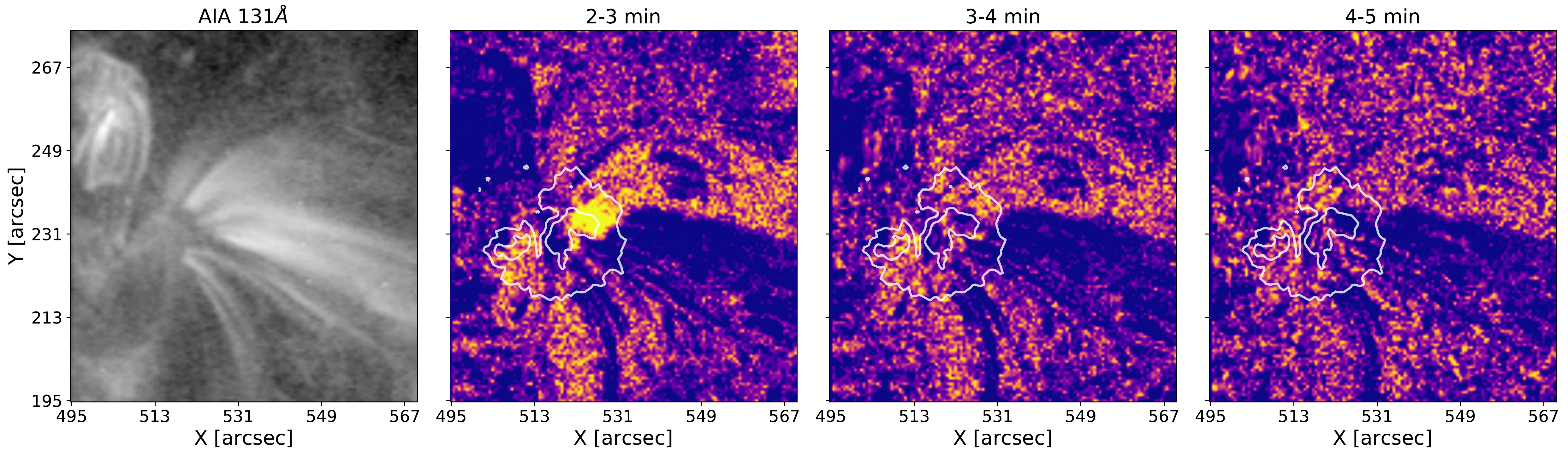}
    \includegraphics[width=0.85\textwidth]{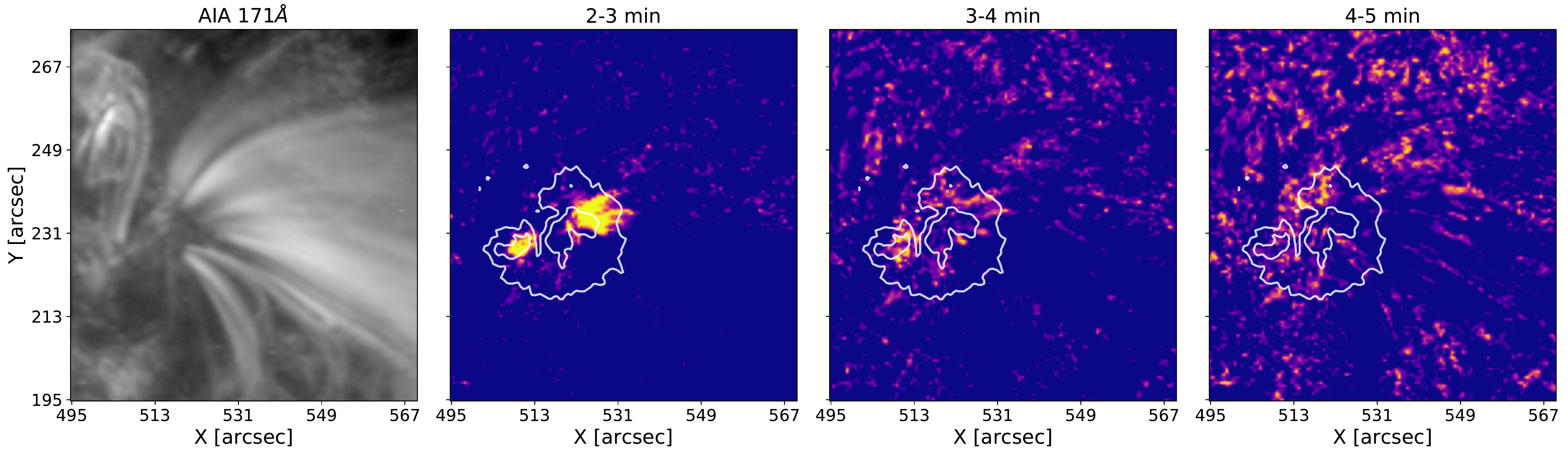}
    \caption{Snapshots of the ROI (first column) taken by SDO/AIA at approx. 00:00 UT for 1700~\AA, 1600~\AA, 304~\AA and 171~\AA (from top to bottom). Maps (from second to fourth column) showing the power distribution obtained from the periodograms calculated for each pixel of the ROI over three 1 minute intervals between 2 and 5 minutes. The second column corresponds to the main powers obtained in the interval [2--3]\,min, the third to [3--4]\,min and the last one to [4--5]\,min. The colour scales in the different columns are normalized to the same value for comparison.}
    \label{fig:pmaps1}
\end{figure*}

\begin{figure*}
    \centering
    \includegraphics[width=0.85\textwidth]{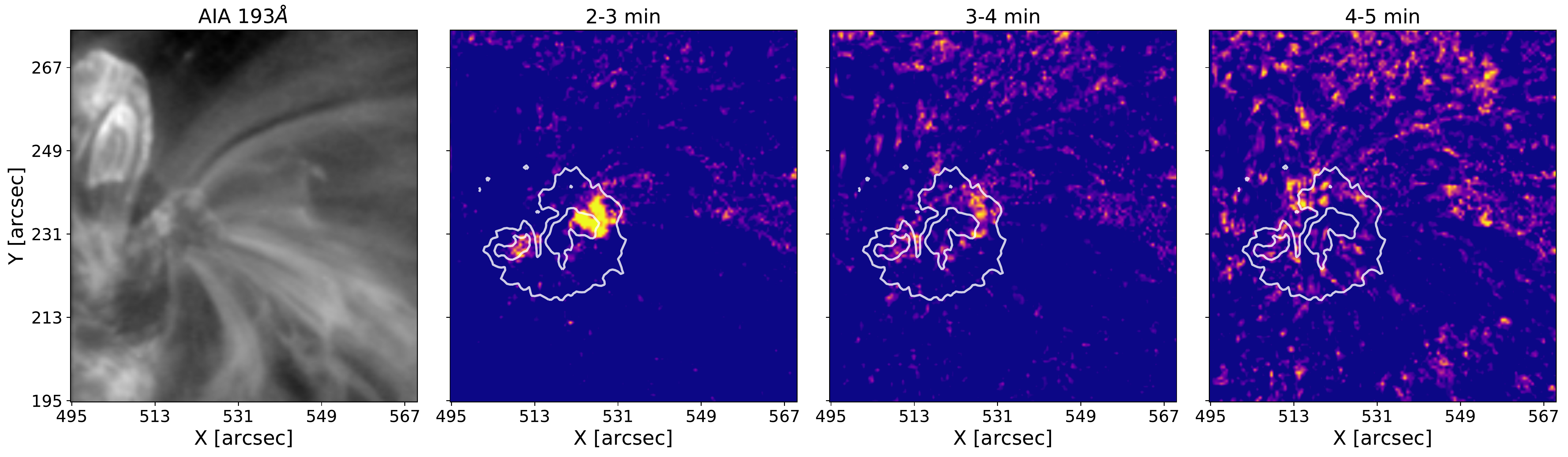}
    \includegraphics[width=0.85\textwidth]{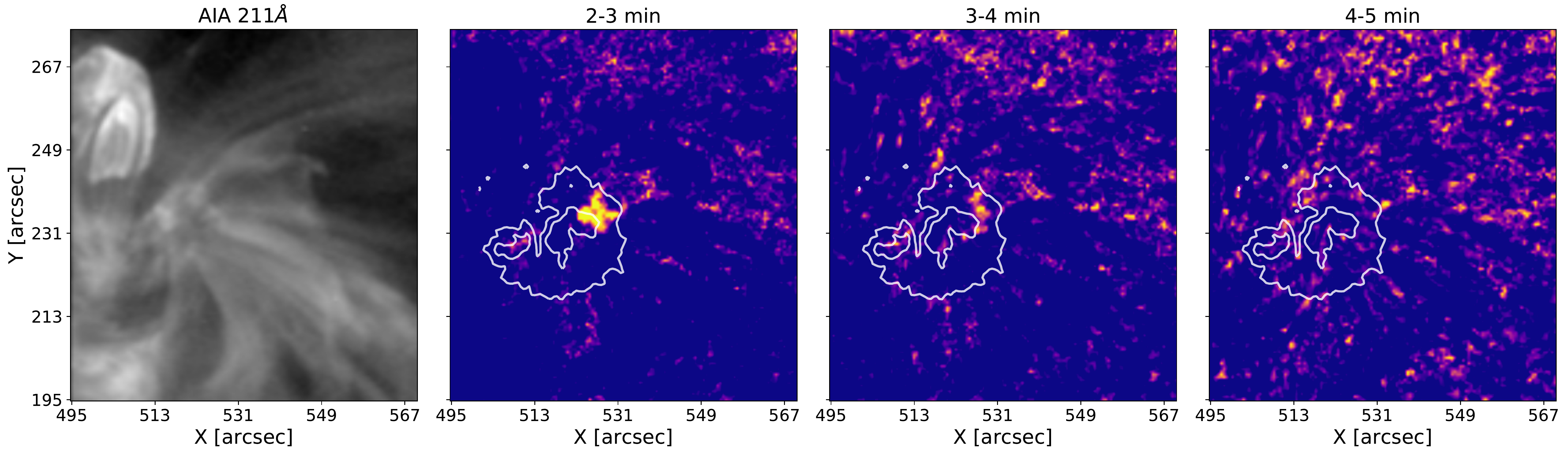}
    \caption{ Same as that of Fig.~\ref{fig:pmaps1} but for wavelengths 193~\AA{} and 211~\AA{}.}
    \label{fig:pmaps2}
\end{figure*}

In order to identify the source and to characterize the observed periodicity of the phenomenon, we track the oscillations throughout the atmosphere, from the photosphere up to the corona. To that aim, we perform a periodicity analysis comprising different heights by computing the Lomb-Scargle periodograms \citep[based on the \textit{fasper} routine described in][]{press2002numerical} of the intensity time series at each pixel of the region shown in Fig.~\ref{fig:roi} (ROI hereafter) considering 120 minutes of observations. From the periodograms obtained, we determine the maximum power value between 1.5 and 15 minutes and normalize the rest of the power spectrum to this value. Then we search for all the local maxima within intervals of 1 minute and choose the largest value among them. In Figs.~\ref{fig:pmaps1}--\ref{fig:pmaps2} (from the second to the last column) we display this value of maximum power for every pixel in each range of periods between 2 and 5 minutes for the different wavelengths (from top to bottom row). The boundaries of the umbra and the penumbra are also shown in these plots. In the first column we show the corresponding observed intensities within the ROI.
%

Note that from the second column of Figs.~\ref{fig:pmaps1}--\ref{fig:pmaps2}, inside the umbra, there is a localized and intense power in  the [2--3] min interval, whose area is growing as the atmospheric height increases (from top to bottom). In the lower atmosphere (1700~\AA{} and 1600~\AA{}) this area is coincident with the footpoints of the fan loops. As the height and the temperature increases i.e., through the 304~\AA{}, 131~\AA{} and 171~\AA{} channels, the area of the power enhancement expands, but a reduction in it is observed in the 193~\AA{} and 211~\AA{} channels. The growth of this area in the [2--3] min power could be attributed to the expansion of the flux tube through which the perturbation travels \citep[see e.g.][]{jess2012}. In contrast, the shrinking of it in the 193~\AA{} and 211~\AA{} channels is likely a temperature effect as these channels observe relatively hotter plasma. It may be noted that the visibility of the associated loop structures is also limited in these channels (see Fig.~\ref{fig:pmaps2}). Surrounding the umbra, the power within the [3--4] min period band (third column) is enhanced, almost as much as that in the [2--3] min band in the photosphere and chromosphere. The area of this enhancement is also expanding outwards with increase in the height but in the corona the power is much less compared to that in the [2--3] min period band. In the lower atmosphere (1700~\AA{} and 1600~\AA{}) the wave power within the [4--5] min band (fourth column) is more intense and ubiquitous compared to the other bands in both the penumbra and quiet region. On the other hand, in the higher atmosphere, especially in the coronal channels, the power within this band does not seem to display any clear enhancements. These results indicate that the oscillations we observe in the corona are localized and have dominant periods between [2--3]\,min. Moreover, they also suggest that these oscillations could have originated in the lower atmosphere from where they propagate up to the corona through an expanding flux tube, that seems to be rooted in the sunspot umbra.  

\subsection{Kinematic Characterization}
\label{ss:kin:s:obs}
\begin{figure}
    \centering
    \includegraphics[width=0.85\columnwidth]{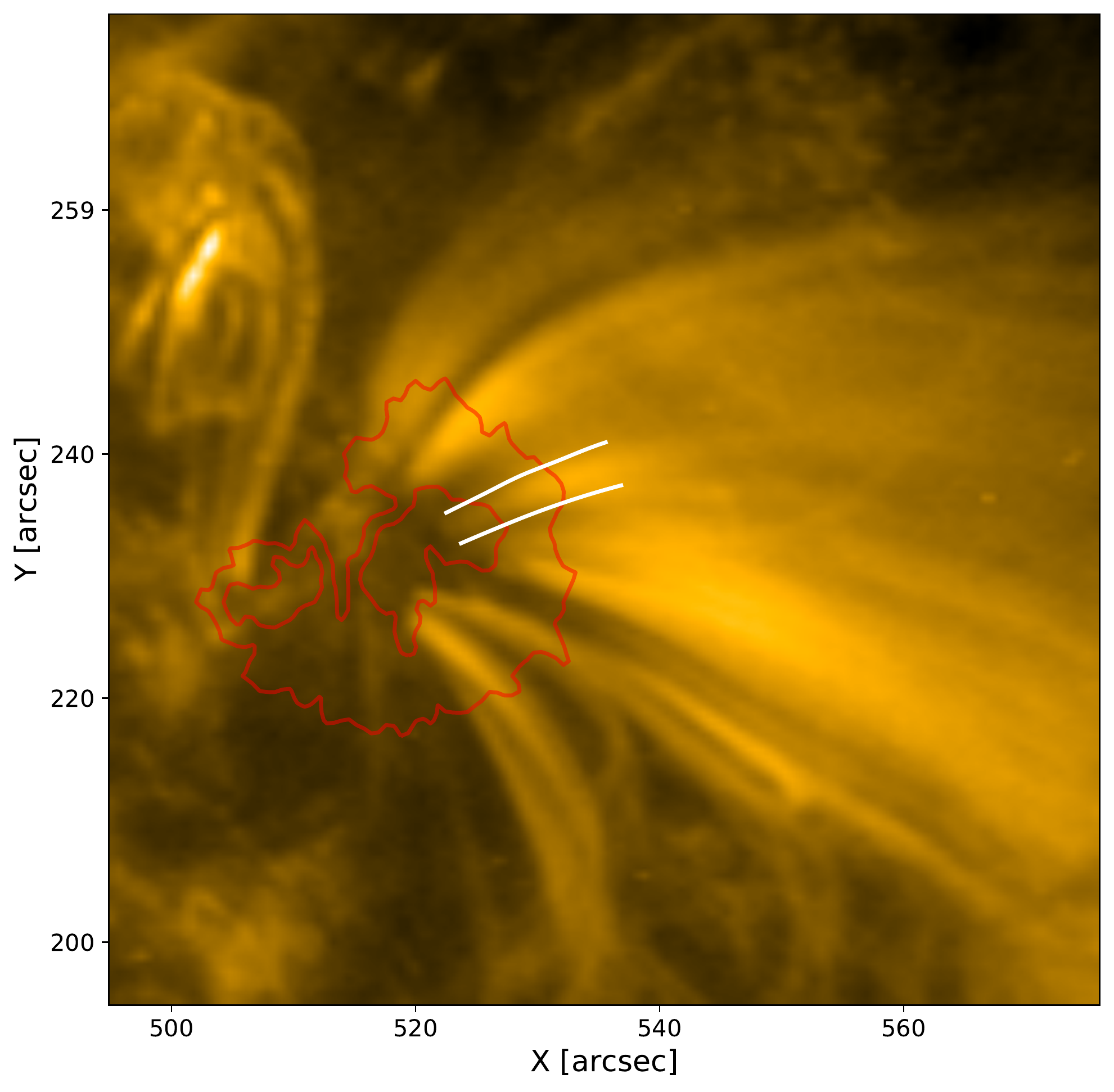}
    \caption{Location of the analyzed loop marked on the 171~\AA{} image. The solid white lines outline the boundaries of the selected loop. The red contours represent the external boundaries of the umbra and penumbra of the sunspot.}
    \label{fig:loop}
\end{figure}

\begin{figure}
    \centering
    \includegraphics[width=\columnwidth]{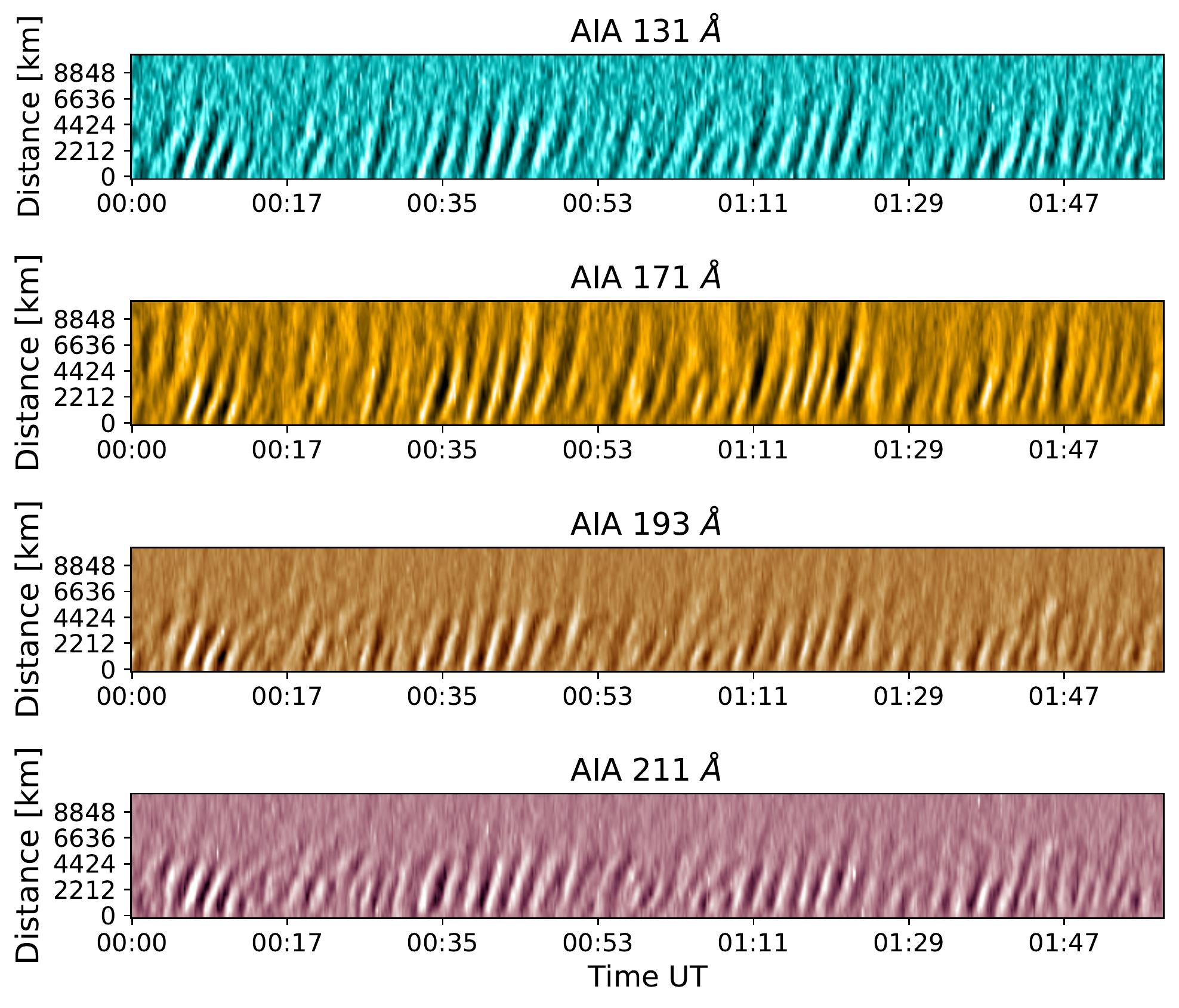}
    \caption{Time-distance ($t-d$) maps obtained for the loop shown in Fig.~\ref{fig:loop} for the 131~\AA{}, 171~\AA{}, 193~\AA{} and 211~\AA{} channels of AIA (from top to bottom).}
    \label{fig:ht_maps}
\end{figure}

To determine the propagation speed of the observed propagating intensity disturbances,
we track the intensity perturbations along a slightly curved slit of finite width located along one of the loops that seem to emanate from the active region. The loop chosen matches the region where the oscillation power is the largest (see Fig.~\ref{fig:pmaps1}). The extent and location of the slit is marked with the white lines in Fig.~\ref{fig:loop}. The longitudinal extension of the loop is $\approx$~10150\,km. The boundaries of both the umbra and the penumbra, as determined from the HMI Continuum image, are also overlaid in this figure with red contours. 
%
By averaging the intensities across the loop we construct time-distance ($t-d$) maps as displayed in Fig.~\ref{fig:ht_maps} for the AIA 131~\AA{}, 171~\AA{}, 193~\AA{} and 211~\AA{} channels. For a detailed description of the method followed, see \citet{krishna2012}. As it can be noted from the figure a regular and recurrent ridge pattern is evident in all the channels. The inclination of the ridges in each $t-d$ map reveals the projected propagation speed of the brightness disturbances in the POS as they propagate along the region defined by the slit. The apparent curvature in the inclination, prominent in 171~\AA{}, where the signal is strong up to the top end of the slit, is a signature of their apparent acceleration. 
%

\begin{figure}
    \centering
    \includegraphics[width=0.8\columnwidth]{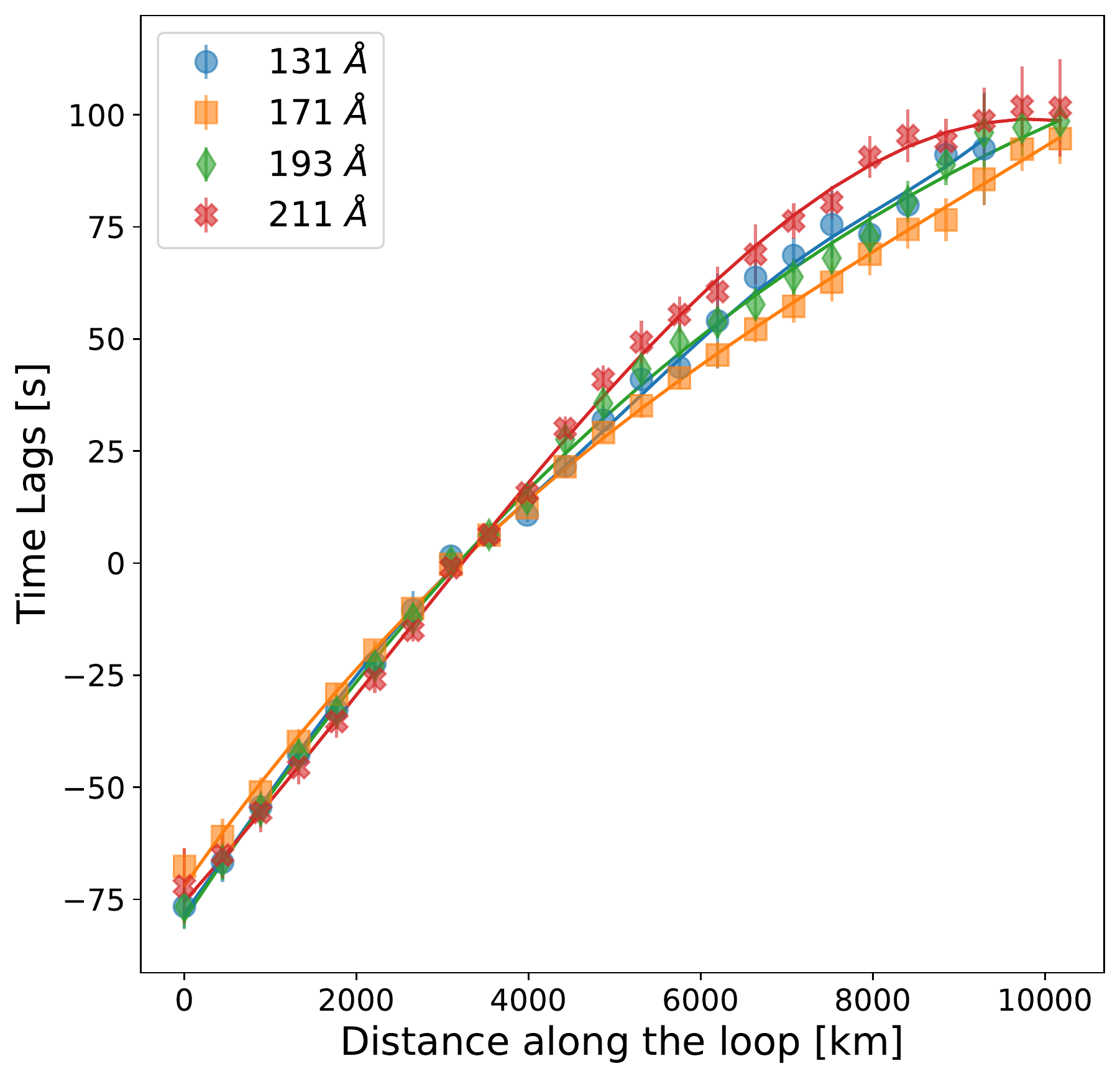}\\
    \includegraphics[width=0.8\columnwidth]{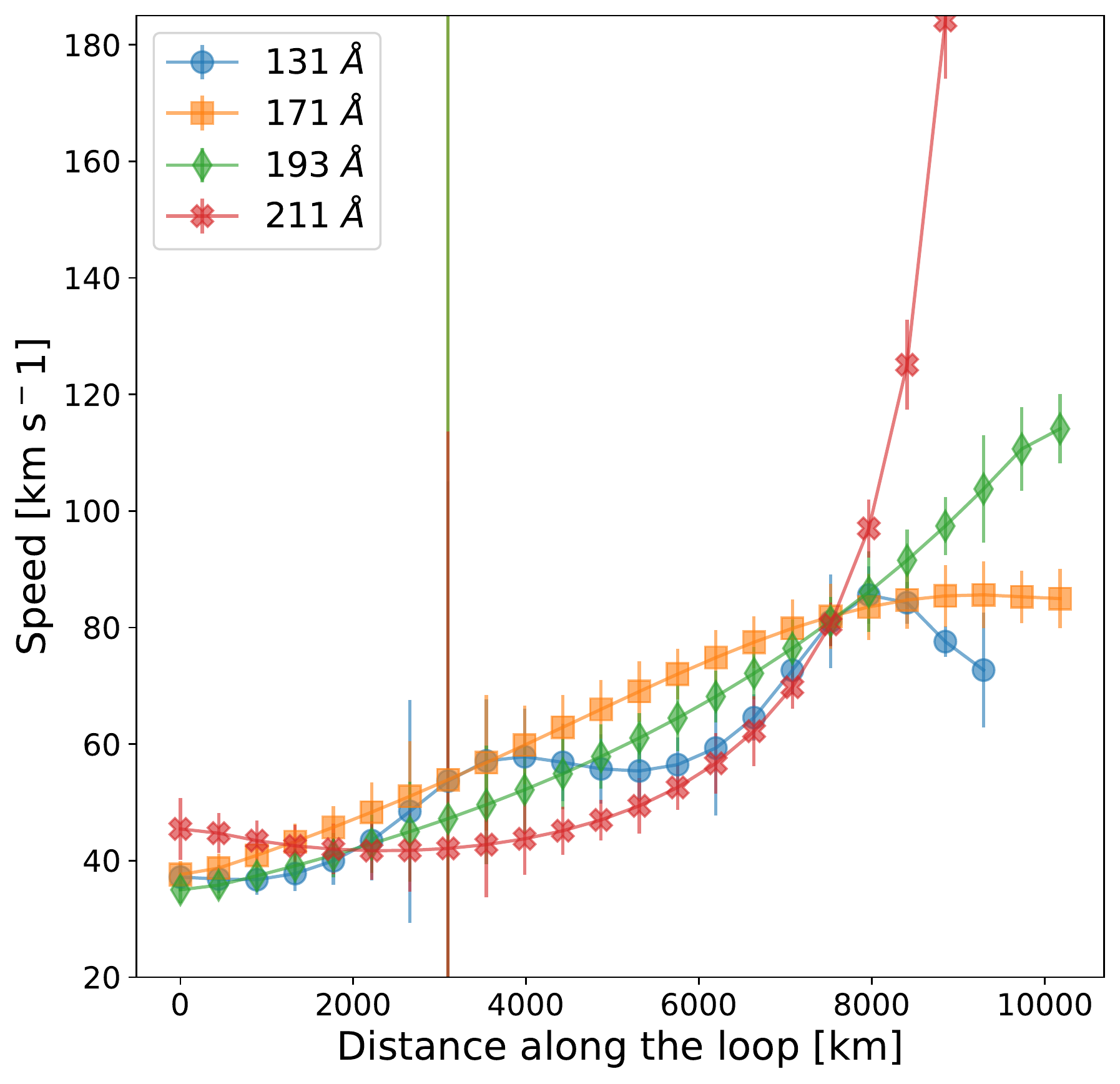}
    \caption{Top panel: Time lags as a function of distance along the loop for the 131~\AA{}, 171~\AA{}, 193~\AA{} and 211~\AA{} channels. The individual values are represented by coloured symbols, the associated errors are shown by the vertical bars and the solid lines denote the corresponding spline fits to the data. Bottom panel: Projected POS speeds derived from the spline fits to the time lag values.}
    \label{fig:speeds}
\end{figure}

To give an account of the kinematics of the phenomenon, a quantitative analysis of the $t-d$ map patterns is required. We thus estimate the POS speeds of the fronts using the method described in \citet{tomczyk2009}. First, for each wavelength, we determine the time lags as a function of distance along the loop. For that purpose, we build a reference light curve from the average intensities across three consecutive rows where the oscillatory pattern is clearly visible. Then we calculate the cross-correlation at all spatial locations with respect to that reference. Additionally, we estimate the errors on the cross-correlation ($\Delta C'$) following equation (13) in \citet{misra2018} that states:
\begin{equation}
    \Delta C'_{XY}=\frac{1-C^2_{XY}}{N^2\sqrt{\sigma^2_X \sigma^2_Y}}\sqrt{\sum^{N/2}_{k=-N/2}|\tilde{X}_k|^2|\tilde{Y}_k|^2}\,,
\end{equation}
where $X$ and $Y$ represent two light curves of $N$ elements, $\sigma_X$ and $\sigma_Y$ are the standard variations of the respective light curves, $C_{XY}=c_{XY}/\sqrt{\sigma_X^2 \sigma_Y^2}$ where $c_{XY}$ is the cross-correlation between the two time series and $\tilde{X_k}$ and $\tilde{Y_k}$ are the discrete Fourier transforms of $X$ and $Y$, respectively, in the frequency domain $k$. Therefore, the uncertainty in the correlation function depends on the value of the correlation and the amplitude of the correlated light curves. For each pair of correlated light curves we take the location of maximum correlation along with the two preceding and the two following points. The correlation values at these locations are then fitted with a parabola function using non-linear least squares method\footnote{We use the \texttt{curve-fit} routine from Scipy optimize package} considering the correlation errors. The vertex of the parabola corresponds to the time lag between that pair. Afterwards we fit a cubic B-spline function\footnote{Using \texttt{splrep} from Scipy interpolate} to the obtained time lags to estimate the POS speeds. The time lags in all the channels and the corresponding fitted curves are shown in Fig.~\ref{fig:speeds} (top panel) as a function of the distance along the loop. The vertical bars on each point represent the respective errors calculated from the uncertainties on the parameters of the parabola fitting. The spline fits shown by the solid lines are weighted by the inverse of these errors. Note that the error in the time lags is minimum near the reference row, at $\approx3600$\,km, because the correlation is maximum there. 

As follows, we estimate the propagation speeds by computing the derivative of the distance along the loop with respect to the fitted time lags. The obtained POS speed values are exhibited in Fig.~\ref{fig:speeds} (bottom panel) as a function of the distance. The different colours represent the values for the different wavelengths and the vertical bars correspond to the propagated errors estimated as $v\, \sigma_{t_{lag}}/|t_{lag}|$, where $v$ is the estimated speed, $t_{lag}$ is the respective time lag and $\sigma_{t_{lag}}$ is the corresponding error. Note that the uncertainties near the reference row are quite large because $t_{lag}\approx 0$. The speeds in all the wavelengths exhibit an accelerated profile in accordance with our visual prediction. Similar results were previously reported by \citet{sheeley2014} and \citet{krishna2017}. Except for a few locations near the top of the loop segment (where the signal is low), the speed values mainly vary from 40 to 100 km\,s$^{-1}$.
It is worth remarking that the exact value of the speed (and in particular of the acceleration) is strongly sensitive to the fitting function used for the time lags (a small change in the fit could result in a larger change in the acceleration profile and hence of the local speed). This may indicate that it is difficult to distinguish the POS speeds in different channels. Nevertheless, we would like to emphasize on the accelerated behaviour but also bear in mind that in 193~\AA{} and 211~\AA{} the signal is low and the loop is not clearly discernible in the upper half part of the slit. Therefore the speeds calculated from these channels above $\approx$5000 km should be considered carefully. 

\section{Numerical model}
\label{s:num}

In order to interpret the accelerated propagation observed in multiple channels, especially to decouple the effect of projection on the propagation speeds, we perform numerical simulations. The simulations were done with the code \mancha \citep[see][]{felipe2010,gonzalez-morales2018,khomenko2018}, which solves non-linear non-ideal equations of the magnetohydrodynamics (MHD) for perturbations. The thermodynamic variables (density $\rho$, gas pressure $p$) and the magnetic field $\vec{B}$ are split into an equilibrium state and their departures from it, while the flow speed, $\vec{v}$, is always treated as (non-linear) perturbation. Consequently, the magnetohydrostatic (MHS) equilibrium is explicitly removed from the equations. This strategy gives certain advantages for the simulations of waves in magnetic structures, since tiny numerical deviations from the MHS equilibrium do not grow in time, unlike in setups that rely on the full variables, guaranteeing a stable initial equilibrium configuration. We can summarise the equations that we used in this work as follows:

\begin{align}
 \frac{\partial \rho_1}{\partial t}&+\nabla \cdot \left[(\rho_0 +\rho_1)\vec{v}_1\right]=\left(\frac{\partial \rho_1}{\partial t}\right)_{\mathrm{diff}} \label{eq:continuity}\\
 \frac{\partial \left[(\rho_0 +\rho_1)\vec{v}_1\right]}{\partial t}&+\nabla \cdot [(\rho_0 +\rho_1)\vec{v}_1\vec{v}^{\mathrm{T}}_1+(p_1+\frac{|\vec{B}_1|^2}{2\mu_0} \label{eq:momentum}\\
&+\frac{\vec{B}_0\cdot \vec{B}_1}{\mu_0})\mathbb{I} -\frac{1}{\mu_0}(\vec{B}_1\vec{B}^{\mathrm{T}}_1+\vec{B}_1\vec{B}^{\mathrm{T}}_0+\vec{B}_0\vec{B}^{\mathrm{T}}_1)] \nonumber \\
&=\rho_1 \vec{g}+\left(\frac{\partial[(\rho_0+\rho_1)\vec{v}_1]}{\partial t}\right)_{\mathrm{diff}} + \vec{S}(t) \nonumber \\
 \frac{\partial e_1}{\partial t}&+\nabla \cdot [(e_0+e_1+p_0+p_1+\frac{|\vec{B}_0+\vec{B}_1|^2}{2\mu_0})\vec{v}_1 \label{eq:energy}\\
 &-\frac{1}{\mu_0}(\vec{B}_0+\vec{B}_1)(\vec{v}_1\cdot (\vec{B}_0+\vec{B}_1))]\nonumber \\
 &=(\rho_0 +\rho_1) \vec{v}_1 \cdot \vec{g}+\left(\frac{\partial \vec{e}_1}{\partial t}\right)_{\mathrm{diff}} \nonumber \\
 \frac{\partial \vec{B}_1}{\partial t}&=\nabla \times [\vec{v}_1 \times (\vec{B}_0 + \vec{B}_1)] + \left(\frac{\partial \vec{B}_1}{\partial t}\right)_{\mathrm{diff}} \label{eq:induction}
\end{align}
where $\vec{g}=-274\,\hat{k}~\mathrm{m\,s^{-2}}$ is the gravitational acceleration in the vertical direction $z$, $\vec{S}(t)$ is a time-dependent external force and $\mathbb{I}$ is the identity matrix. The subscripts ``0'' and ``1'' indicate the background variables and their perturbations, respectively. The background state should satisfy the MHS equilibrium. The total energy per unit volume is defined as $e=\frac{1}{2}\rho v^2+\frac{p}{\gamma -1}+\frac{B^2}{2\mu_0}$ and satisfies that $e=e_0+e_1$.
The terms subscripted with “diff” are numerical diffusion terms required for numerical stability and are computed according to \citet{vogler2005} and \citet{felipe2010}.

In the present numerical study we perform 2D simulations on a Cartesian grid of 1344~x~672 cells distributed over a domain of $[4\times 10^4] $~x~$[2 \times 10^4]$ km in the horizontal ($x$) and vertical ($z$) directions, respectively. This results in a resolution of $\approx$30 km. The horizontal coordinate varies from $-2\times 10^4$  to $2\times 10^4$ km and the vertical coordinate from 0 to $2\times 10^4$ km, where $0$ km coincides with the base of the photosphere. 

\subsection{Magnetohydrostatic equilibrium}
\label{ss:mhs:s:num}
\begin{figure*}[t]
    \centering
    \includegraphics[height=0.55\columnwidth]{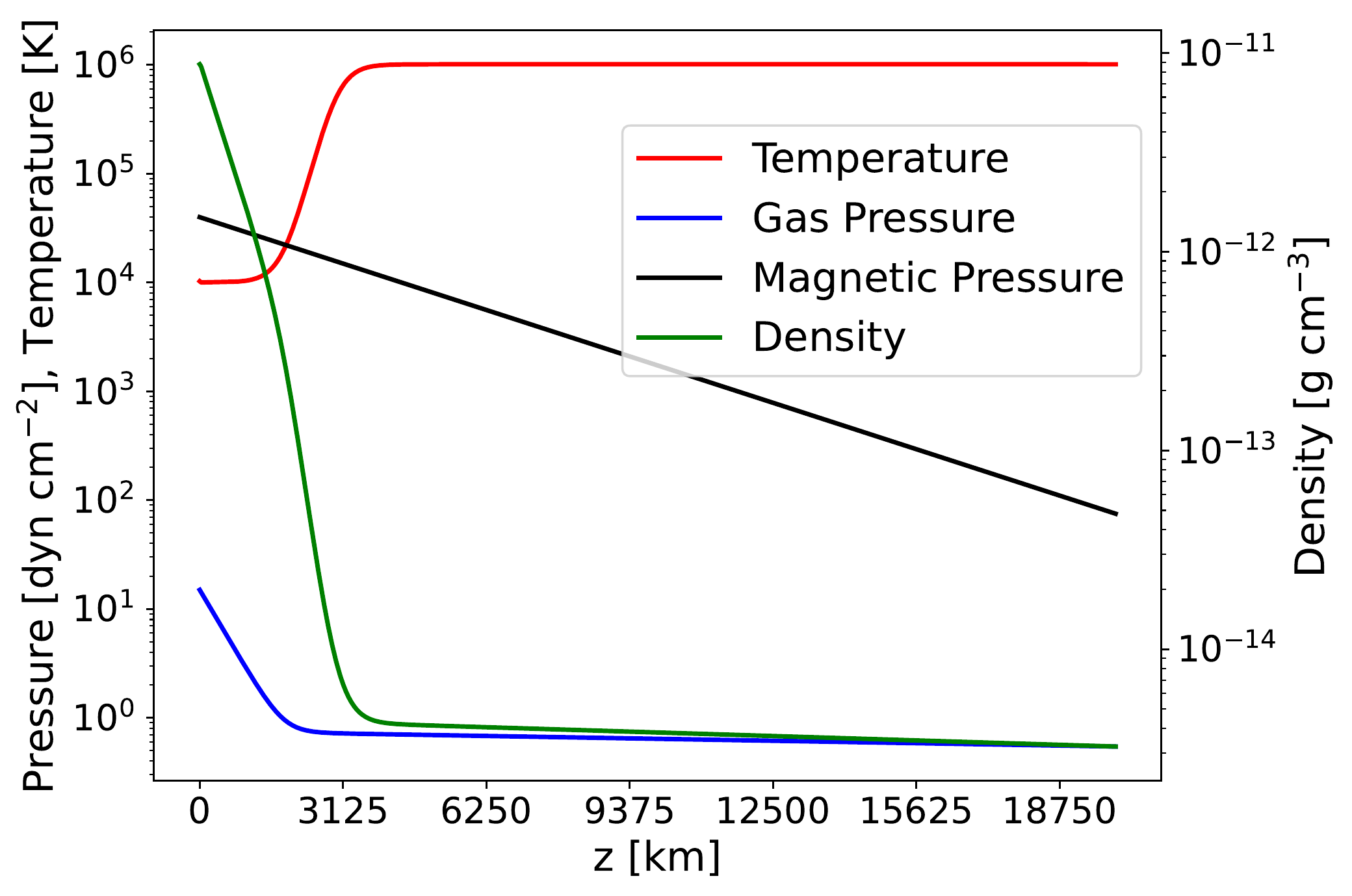}
    \includegraphics[height=0.55\columnwidth]{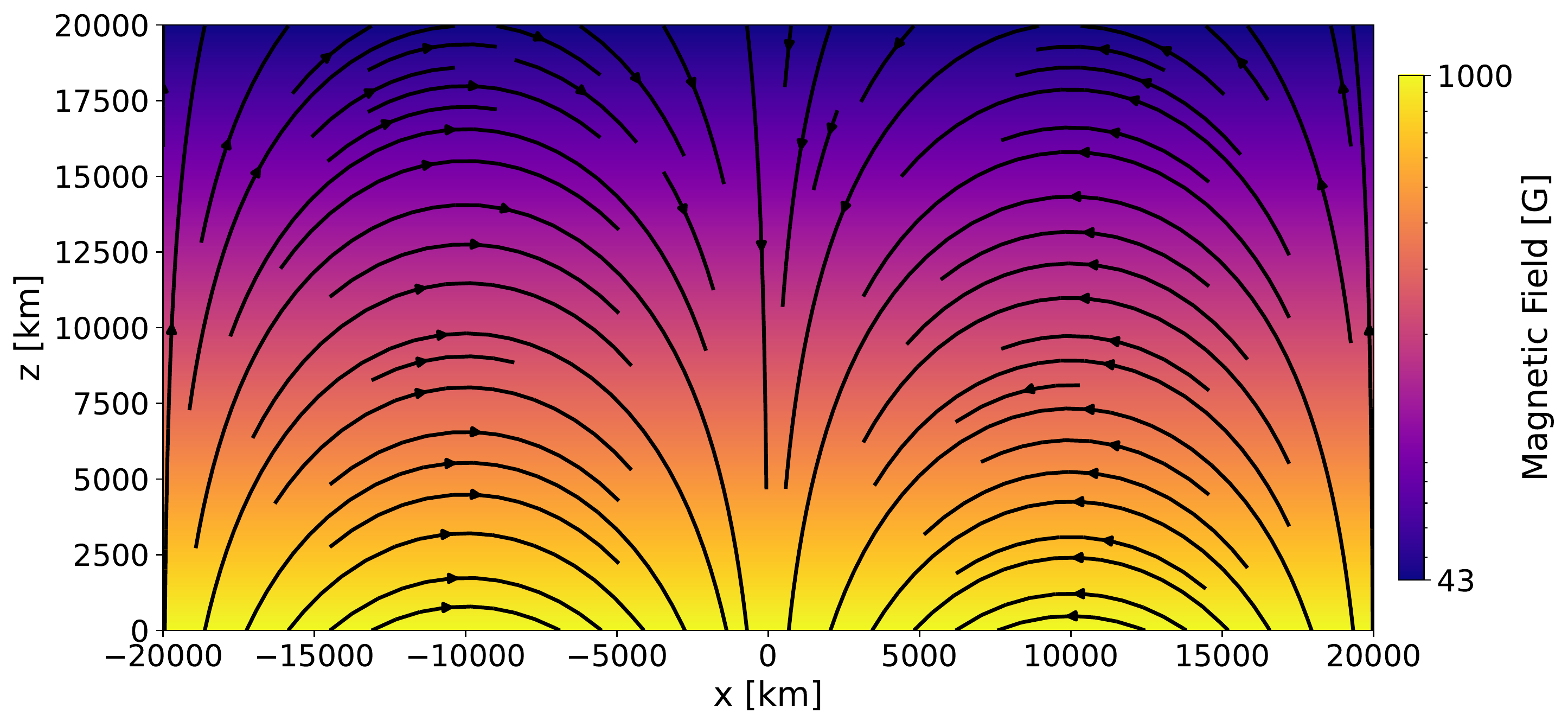}
    \caption{Initial conditions for the simulation. Left panel: Temperature (red  line), density (green line), gas (blue line) and magnetic pressure (black line) stratification along the $z$ direction considering a constant value of 274 m\,s$^{-2}$ for gravity. Right panel: Symmetric potential magnetic field configuration. The background colour denotes the intensity of the magnetic field while the solid black lines show some representative magnetic field lines.}
    \label{fig:mhs}
\end{figure*}
As we mentioned before, the \mancha code ensures the magnetohydrostatic equilibrium of both stratified atmosphere and magnetic configuration when the initial condition is in such state. In order to have a background equilibrium we need to satisfy $-\nabla p_0 + \rho_0 \vec{g} + \vec{j_0}\times \vec{B_0}=0$, where $\vec{j_0}$ is the current density (other variables have been already defined). To model the initial atmosphere stratification we choose a smooth temperature profile given by:
\begin{equation}
   T_0(z)=a_0 \tanh\left(\frac{z-a_1}{a_2}\right)+a_3
   \label{eq:temp}
\end{equation}
where $a_0=5 \times 10^5$ K, $a_1=3 \times 10^3$ km, $a_2=5 \times 10^2$ km, and $a_3=5.1 \times 10^5$ K. While the parameters $a_{0,3}$ constrain the maximum and minimum value of the temperature, $a_1$ controls the height at which the transition region begins and $a_2$ regulates the ``sharpness'' of it. As a result of the combination of the parameters we obtain a minimum temperature of $10^4$ K in the lower atmosphere and $10^6$ K in the corona. This profile can resemble the thermal inversion in the corona and it was previously used, for example, in the analytical and observational work by \citet{zurbriggen2020} where the authors studied how the different stratifications and magnetic field can influence the cut-off periods of magnetoacoustic-gravity waves. Once we have established the temperature stratification, we obtain the pressure and the density profiles from the hydrostatic equilibrium equation:
\begin{equation}
  \nabla p_0=\rho_0 \vec{g}
  \label{eq:hydrostatic}
\end{equation}
accompanied by the ideal gas law with constant mean molecular weight of $\mu=0.5$. A value of $1.5\times 10^{1}$ dyn\,cm$^{-2}$ was taken for the gas pressure at the base of the atmosphere. Since the stratification is along the vertical direction, equation \ref{eq:hydrostatic} becomes a unidimensional equation along $z$ and in the horizontal direction $x$, the thermodynamic variables are homogeneous. We show the initial temperature, density and pressure in Fig.~\ref{fig:mhs} (left panel) as a function of $z$ in the middle of the domain. 

Furthermore, to ensure the MHS equilibrium, i.e. to satisfy the force equation, we consider a force-free ($\vec{j_0}\times \vec{B_0}=0$) magnetic field as follows:
\begin{align}
    B_{x0}(x,z)&=-B_{\rm base}\exp(-k_z z)\sin(k_z x) \label{eq:magx}\\
    B_{z0}(x,z)&=-B_{\rm base}\exp(-k_z z)\cos(k_z x) \label{eq:magy}
\end{align}
where $k_z=\pi/L$ is the wave number, $L=2\times 10^4$ km, and $B_{\rm base}=1 \times 10^{3}$ G is the value of the magnetic field in the lower atmosphere. The magnetic field, displayed in Fig.~\ref{fig:mhs} (right panel), is symmetric with respect to the central vertical axis and it decreases exponentially with height, resulting in a vanishing field at a large distance. This simple model represents a first order approximation of a sunspot \citep{aschwanden2004} and it is similar to that used in \cite{santamaria2015} (without the null point) and to the extrapolated magnetic field in \cite{jess2013} derived from the observations.

\subsection{Wave generation}
\label{ss:driver:s:num}
\begin{figure}
    \centering
    \includegraphics[width=\columnwidth]{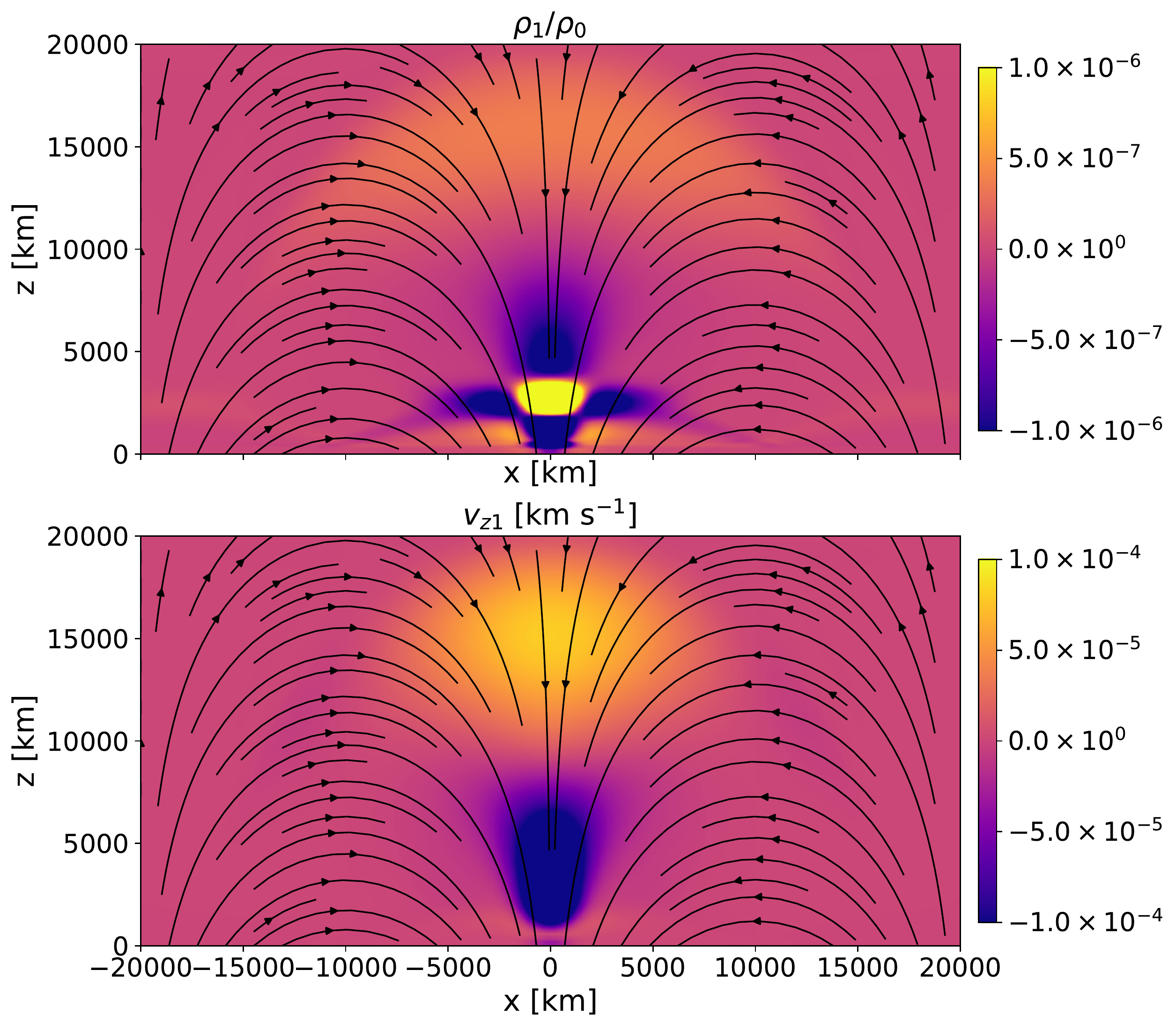}
    \caption{Snapshots of the simulation at $t=198$ s. The upper panel shows the perturbation of the density over the background density ($\rho_1/\rho_0$) and the bottom panel the vertical velocity ($v_{z1}$). The solid black lines represent some of the background magnetic field lines.}
    \label{fig:rho1_v1}
\end{figure}

Based on the oscillation properties described in section \ref{ss:periods:s:obs}, we have a localised region with dominant periods in the [2--3] min interval and the oscillations in the corona appear to be generated in the lower atmosphere. Therefore we employed a source $\vec{S}(t)$ in the vertical direction of the momentum equation (\ref{eq:momentum}) to drive the waves. This source acts as an external force in a localized region close to the bottom boundary of the simulation domain, namely, the lower atmosphere. The source term depends on several free parameters that could vary during the simulations. In the current case we used the following expression: 
\begin{equation}
\vec{S}(t)=A\rho_0 \vec{g} \exp\left(-\frac{(x-x_0)^2}{\sigma_x^2} -\frac{(z-z_0)^2}{\sigma_z^2}\right) \sin(\omega t)
\end{equation}
where $A=5\times 10^{-5}$, 
$x_0=0$, $z_0=20\times dz\approx 595~\mathrm{km}$, $\sigma_x=33\times dx\approx 982~\mathrm{km}$, $\sigma_z=3\times dz \approx 89~\mathrm{km}$, and $\omega=2\pi/180~\mathrm{s^{-1}}$, with $dz=dx\approx30$ km. The source produces self-consistently the perturbations in the velocity and other thermodynamic variables besides the magnetic field. In Fig.~\ref{fig:rho1_v1} we show the variation of the density (top panel) and the velocity (bottom panel) at $t=198$~s. 
Since the source is located at the axis of the magnetic structure where the magnetic field is essentially vertical, it produces slow magneto-acoustic waves as one would expect in a low plasma-$\beta$ atmosphere. These waves propagate along the field (the magnetic field lines are indicated as black lines in Fig.~\ref{fig:rho1_v1}) through the chromosphere, to the transition region, and to the corona, albeit with a significant reflection from the transition region due to the sharp gradient in the sound speed.

It is important to mention that we kept the source amplitude very low in the simulations (though fully non-linear equations were solved), since we aimed at reproducing a linear wave behaviour. Observations do not show significant evidence for nonlinear effects in the propagation of slow waves along coronal loops. Non-linearities affect the wave propagation speed, possibly making it larger than the local sound speed. By keeping the driver amplitude low, we avoid this nonlinear effect in our simulations. 

The boundary conditions on the horizontal direction are periodic. For the vertical boundaries, both up and down, we use a sponge layer to absorb the coming waves. While \mancha has a perfectly matched layer (PML) boundary condition \citep{berenger1994,berenger1996,hu1996,hu2001,parchevsky2007,parchevsky2009}, we found it did not work well for the extremely long-wavelength perturbations in the corona. Instead, a simple sponge layer was found to be a better choice \citep[see][]{gonzalez-morales2019}. For the sponge we used 10 grid points at the bottom and 60 at the top.

\section{Numerical Results}
\label{s:results}
\begin{figure}
    \centering
    \includegraphics[scale=0.4]{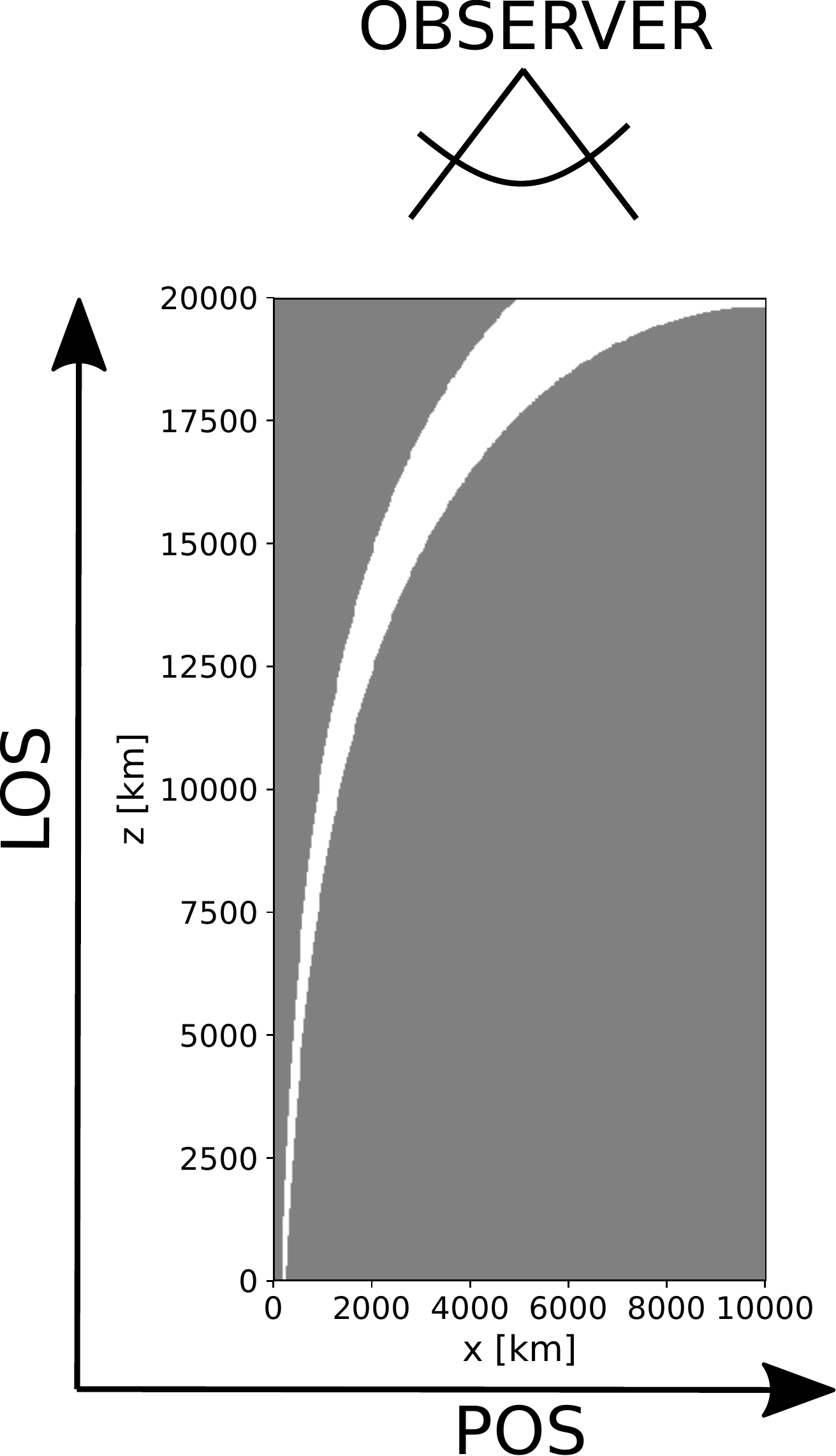}
    \caption{Location of the flux tube extracted from within the simulation domain. The external boundaries are defined by two field lines from the magnetic field equations (\ref{eq:magx}--\ref{eq:magy}). The LOS is aligned with the $z$-axis of the numerical domain and the POS with the $x$-axis. The observer location is displayed at the top of the domain.}
    \label{fig:fluxtube}
\end{figure}

In order to facilitate a direct comparison with the observations we generate synthetic images in the EUV spectrum from the simulation results. Since the oscillation amplitudes in the simulation are quite low, we first re-scaled the perturbations in all the quantities by a factor of 10. This scaling  is allowed for perturbations in a linear regime, and brings the oscillation amplitudes close to the observed values. After that we extract a 2D flux tube from the whole numerical domain to emulate one of the fan loops observed in the EUV channels (see Fig.~\ref{fig:roi}). The flux tube, shown in Fig.~\ref{fig:fluxtube}, is defined by all the plasma contained between two magnetic field lines defined by equations (\ref{eq:magx}--\ref{eq:magy}).

\begin{figure}
    \centering
    \includegraphics[width=\columnwidth]{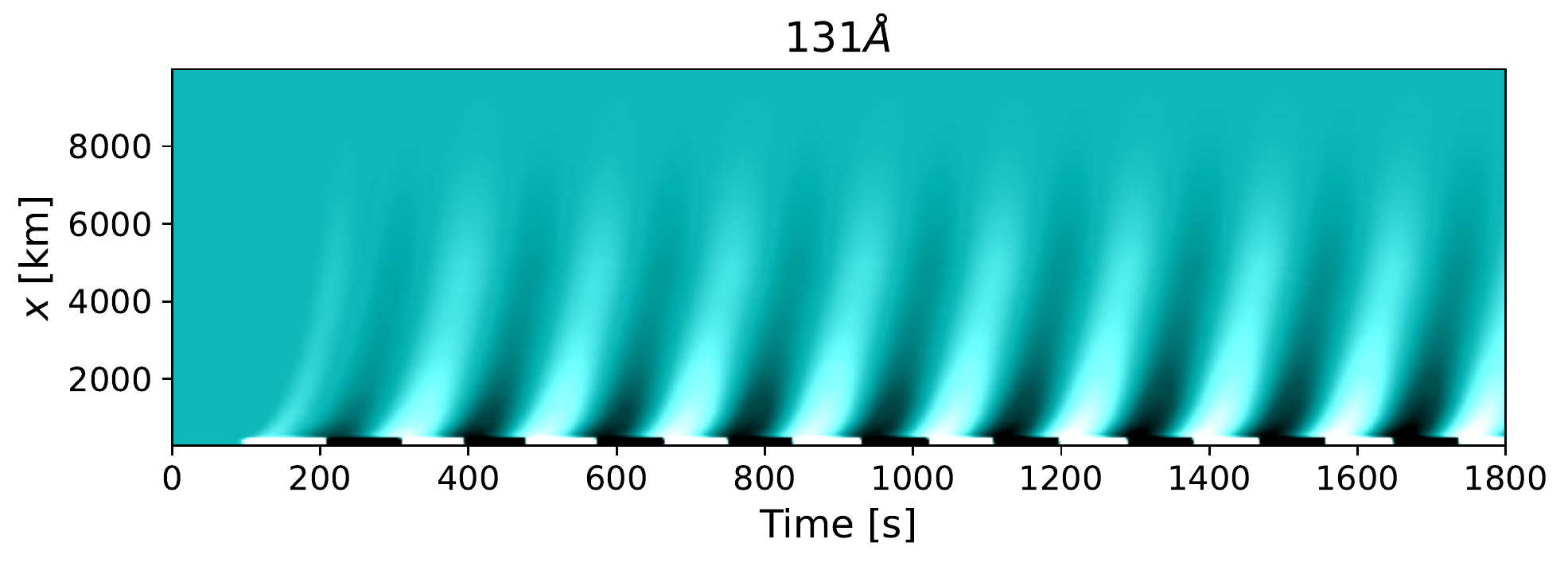}
    \includegraphics[width=\columnwidth]{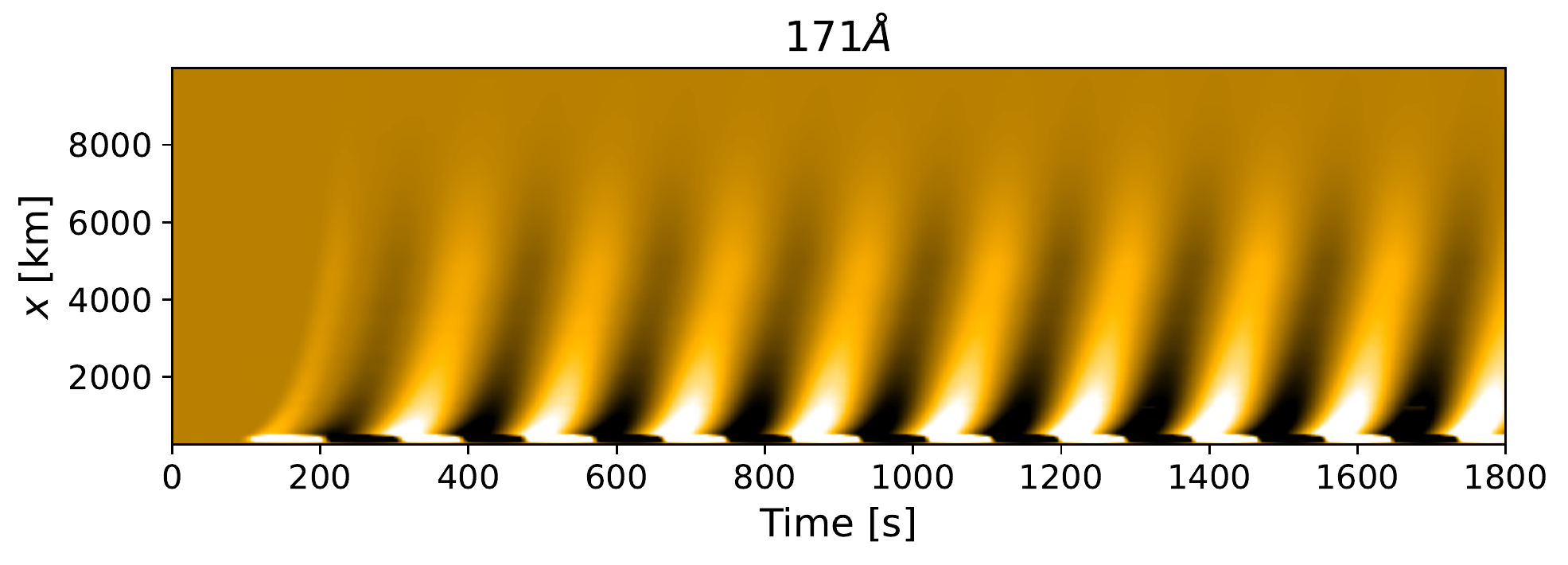}
    \includegraphics[width=\columnwidth]{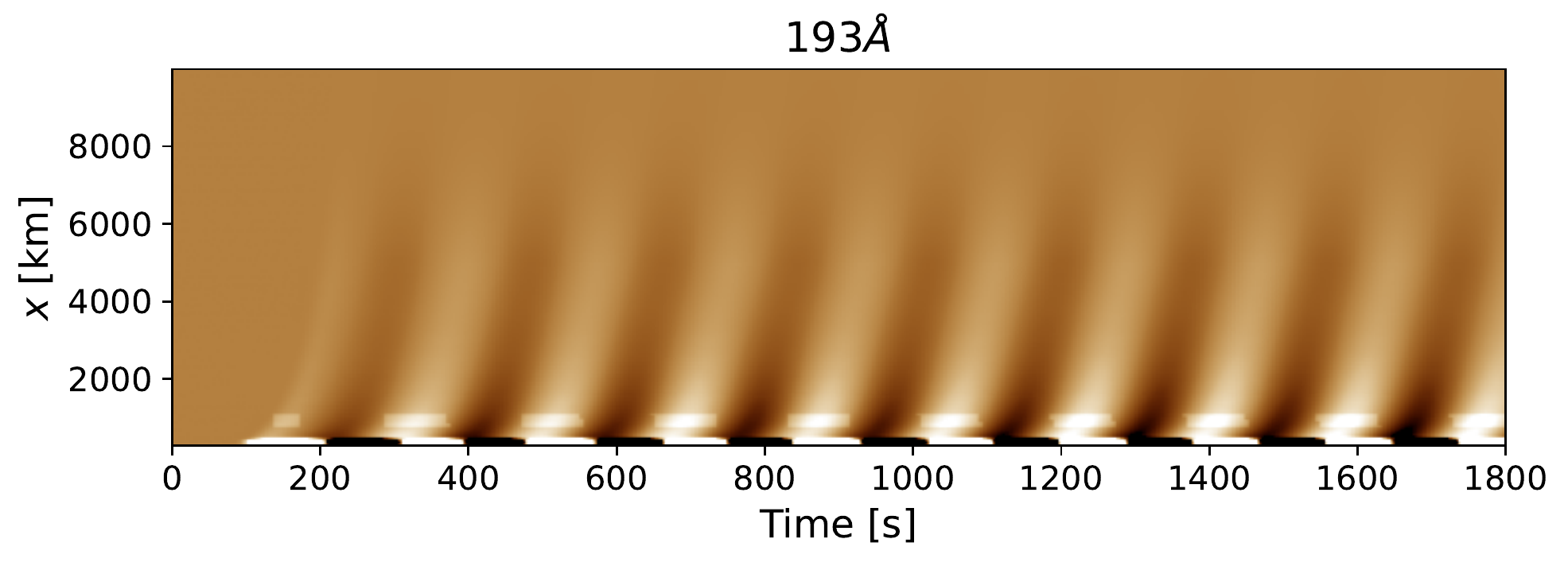}
    \includegraphics[width=\columnwidth]{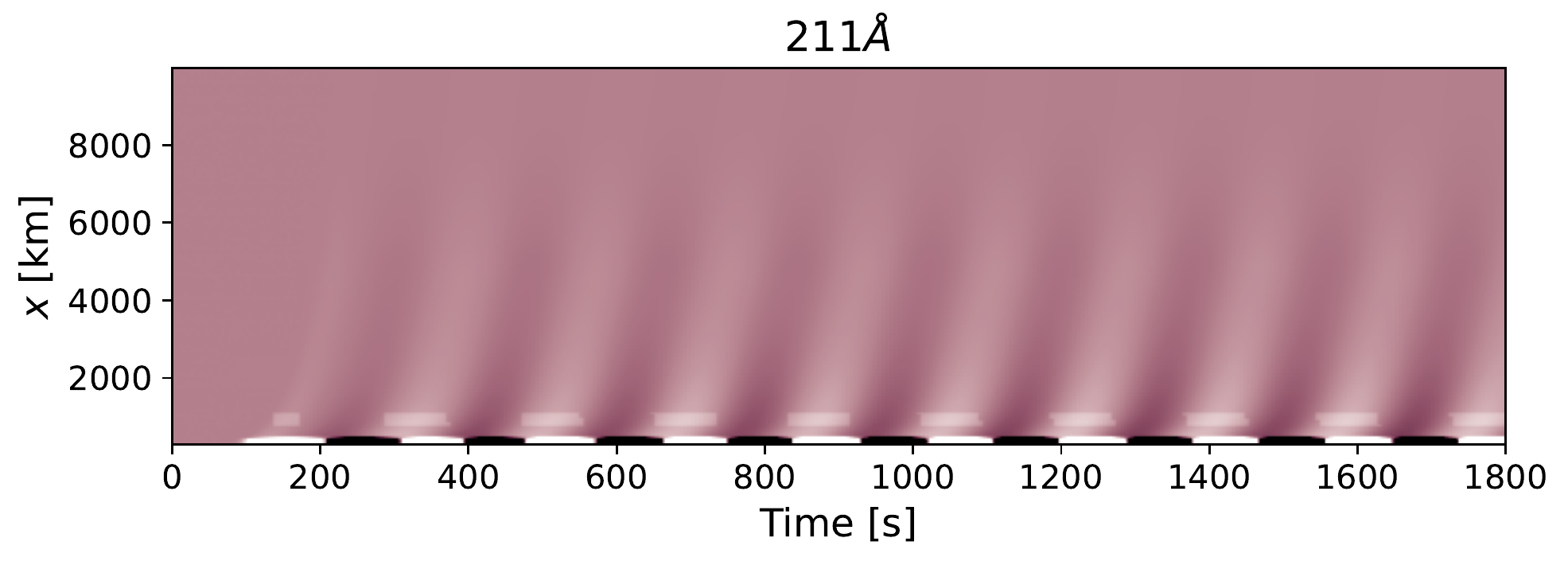}
    \caption{Synthetic time-distance maps ($t-d$) in the 131~\AA{}, 171~\AA{}, 193~\AA{} and 211~\AA{} channels (from top to bottom) obtained from the forward modelled intensities along the selected flux tube considering a projection direction parallel to the $x$-axis as shown in Fig.~\ref{fig:fluxtube}.}
    \label{fig:ht_maps_simu}
\end{figure}

Once we extracted the flux tube, we compute the specific intensity in the EUV spectral lines using the Leuven Forward Modelling code for coronal emission \citep[FoMo,][]{vandoorsselaere2016}. Briefly, given the numerical density, temperature and velocity, this code calculates the specific intensity for a certain monochromatic spectral line in the POS integrating the emissivity along the LOS. The POS together with the LOS are not necessarily aligned with the axes of the numerical simulations. Therefore, for generating the synthetic images, it is possible to choose the spectral line from a specific telescope and to give different viewing angles. In this study, we choose to forward model the intensity for the four EUV  wavelength channels, the 131~\AA{}, 171~\AA{}, 193~\AA{} and 211~\AA{} of AIA. We assumed that the LOS is aligned with the $z$-axis of the numerical domain and one of the axes of the POS is parallel to the $x$-axis. In terms of FoMo parameters this implies  that the viewing angles selected in the code are (0,0), thus the observer is located perpendicularly above of the simulation domain (see Fig.~\ref{fig:fluxtube}). Then we obtain the intensity $t-d$ maps in the four wavelengths, as presented in Fig.~\ref{fig:ht_maps_simu}. As can be seen, all these maps exhibit an accelerated recurrent pattern with no significant differences between the wavelengths. This is due to the narrow temperature distribution of the isolated fluxtube (around $1 \times 10^6$ K) that we considered for calculating the synthetic intensity.  

\begin{figure}
    \centering
    \includegraphics[width=0.8\columnwidth]{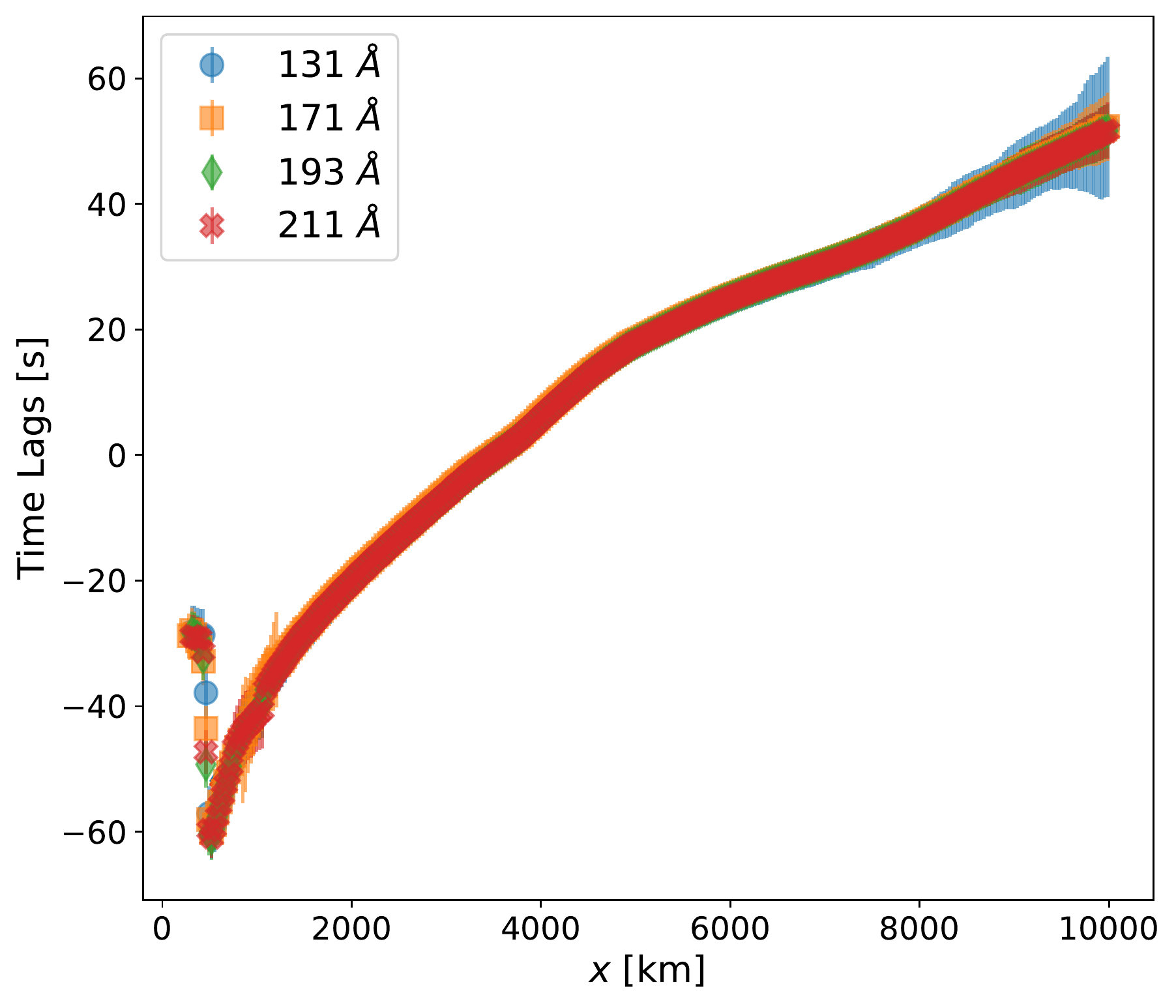}
    \includegraphics[width=0.8\columnwidth]{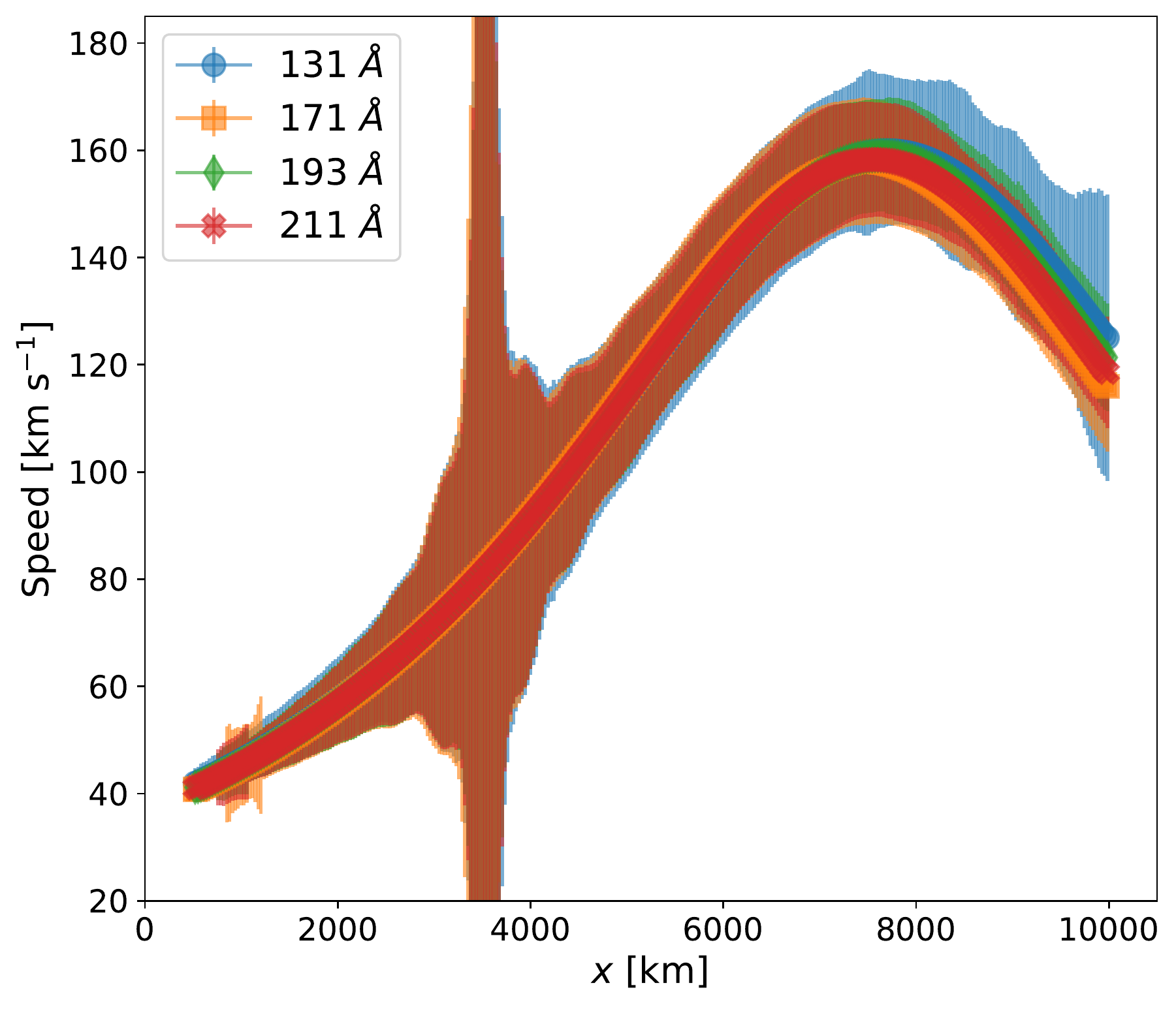}
    \caption{Top panel: Time lags (coloured symbols) obtained from the synthetic $t-d$ maps for each wavelength as a function of the projected distance along the loop. Vertical bars (comparable to the size of the symbols) on these values indicate the associated errors from the quadratic fit performed to obtain the time lags. The solid lines denote spline fits to the data, which are almost indistinguishable. Bottom panel: Propagation speeds obtained from the spline fits to time lag values along with the associated errors.}
    \label{fig:speeds_simu}
\end{figure}

To study the propagation speeds in each spectral line we first determine the time lags as a function of distance in the same way that we calculated for observations. The only difference here is that the reference light curve is selected from a single row rather than averaging three rows, as these data are not noisy. In Fig.~\ref{fig:speeds_simu} (top panel) we show the time lags computed for each passband as a function of the projected distance along the $x$-axis, together with the fitted spline curves. The different coloured symbols represent the values obtained for each wavelength and the vertical lines on them show the associated errors calculated in the same way as that for observations. Note that the error bars here are smaller (comparable to the size of the symbol) than in observations mainly because the correlation values are higher. The solid lines mark the respective spline fits to the data but they are not discernible due to the high density of points (because of the high resolution of simulations). Subsequently, taking the time derivative on the distance using the fitted time lag values, we determine the propagation speeds for each spectral line. We display the obtained values along with their respective errors in Fig.~\ref{fig:speeds_simu} (bottom panel). The errors are again large near the reference row ($\approx$3600\,km) as the time lags are close to zero and the propagated errors are inversely proportional to them. The speed values range from 40 to 160 km\,s$^{-1}$ and represent the apparent propagation speeds seen by an observer located above the simulated active region. They reach a maximum value quite close to the sound speed at about 7300 km distance from the base of the flux tube. Note, however, that the distance is given in the projected coordinate i.e., in the $x$-direction, and, as can be seen from Fig.~\ref{fig:fluxtube}, the flux tube already reaches the top of the domain by this location. Therefore, the apparent speed beyond this location does not correspond to the upward propagation of waves but rather represents the speed at which the waves arrive at the adjacent locations at the top of the domain. This explains why the apparent propagation speed decreases at distances larger than 7300\,km. In addition, the time lags obtained for different wavelengths are almost the same and, therefore, there is no distinction in speeds between them. However, the acceleration dependence on the distance is evident.   

\section{Discussion}
\label{s:discussion}

The observed apparent speeds calculated in Sect.~\ref{ss:kin:s:obs} for the 131~\AA{}, 171~\AA{}, 193~\AA{} and 211~\AA{} passbands vary from 40 to 100 km\,s$^{-1}$ and exhibit accelerated profiles. These values are smaller than the slow magnetosonic wave speeds calculated from the characteristic temperature of each channel. 
Interpreting the observed oscillations as slow waves that are generated in the lower atmosphere and propagate upwards following the magnetic field (see analysis from Sect.~\ref{ss:periods:s:obs}), their accelerated profiles may imply an increase in the local plasma temperature with distance. However, a changing inclination of the magnetic field with respect to the line-of-sight would result in a similar behaviour because of the projection effect. Moreover, this would also explain why the speed values are lower compared to the sound speed. Therefore a combination of both, the physical and the non-physical effects, is possible in observations. The observed speed values are similar to the 171~\AA{} passband measurements described by \citet{krishna2017} and to the sunspot waves events analysed by \citet{sheeley2014}. But, in contrast to \citet{krishna2017}, there is no clear distinction between the speeds determined in 131~\AA{} and 171~\AA{} channels. This suggests that the differences we find in the observed speeds across different wavelengths perhaps do not indicate a multi-thermal structure of the coronal loop. However this is difficult to determine conclusively, as the speeds are subject to large uncertainties that depend on the method and the differences were marginal. In any case, the absence of a clear temperature dependence in the propagation speeds and the diminished visibility of the loop itself (see Fig.~\ref{fig:pmaps2}) in the hotter channels (193~\AA{}, and 211~\AA{}) imply a narrow temperature distribution within the loop.

\begin{figure}
    \centering
    \includegraphics[width=0.8\columnwidth]{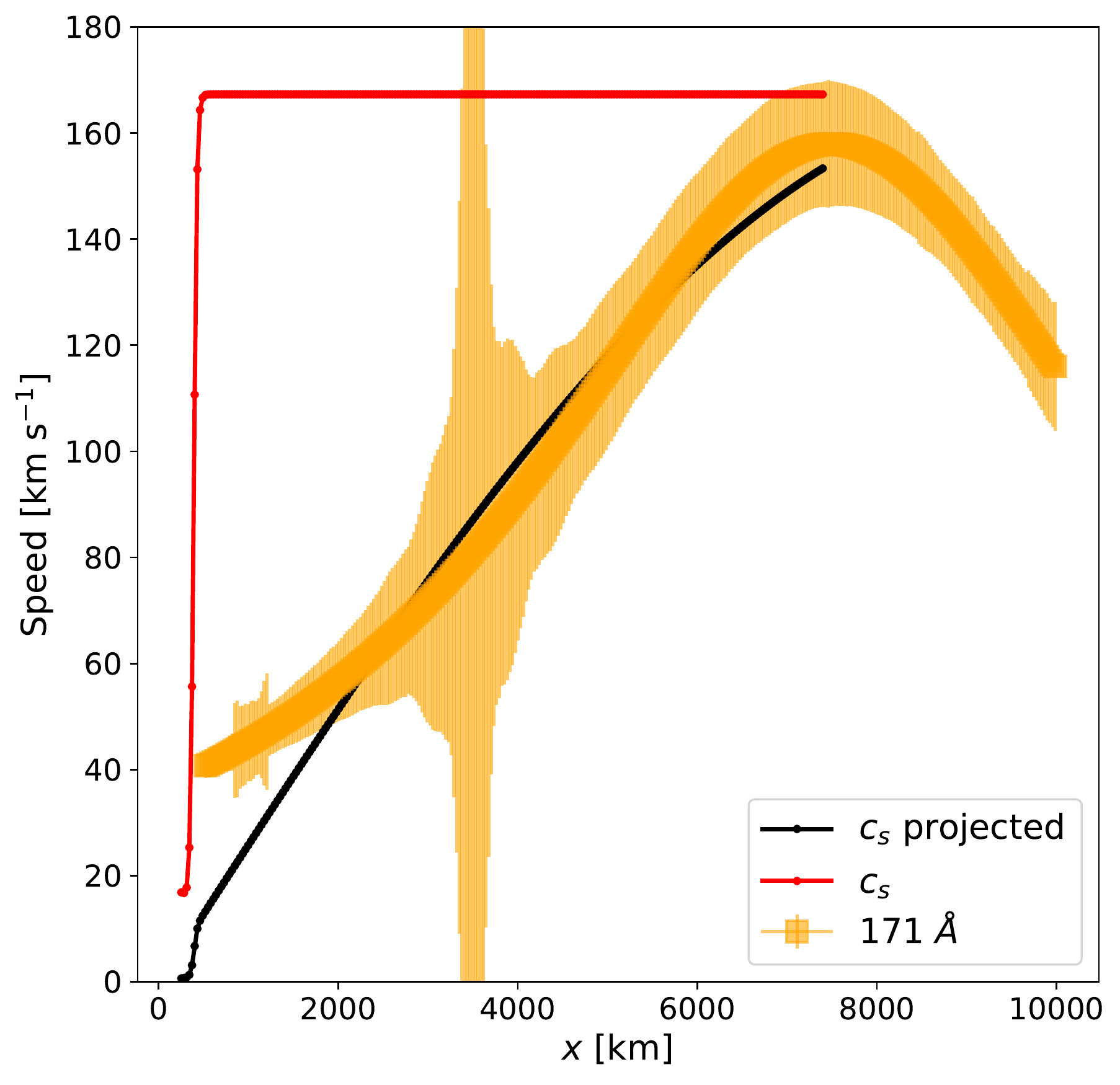}
    \caption{Propagation speed for 171~\AA{} (orange symbols), background sound speed $c_s$ (red symbols) and its projection (black symbols) considering the inclination of the magnetic field as a function of the projected distance along the loop.}
    \label{fig:cs-proj}
\end{figure}

In our simulations, we excite slow waves that propagate parallel to the field lines at the sound speed $c_s$. To demonstrate this we display in Fig.~\ref{fig:cs-proj} the background sound speed $c_s$ (red symbols) for all the locations along the middle field line of the flux tube together with its projection (in black symbols) considering the inclination of the magnetic field with respect to the LOS. The speed calculated from the synthetic $t-d$ map for 171~\AA{} is also shown for reference.
Note that for the coronal conditions employed in the model, $c_S=167~\mathrm{km\,s^{-1}}$  (assuming a mean molecular weight $\mu=0.5$, for a temperature of $1\times 10^6$ K). 
Since there is hardly any variation in temperature in the coronal part of the model (see from the figure that $c_s$ is constant for the locations that correspond to the flux tube), the lower (than $c_s$) propagation speeds along with the accelerated pattern obtained from the synthetic $t-d$ map highlight the projection effect caused by the inclination of the waveguide with respect to the LOS. Indeed, the speed calculated in 171~\AA{} is comparable to the projected sound speed at most of the locations. The discrepancy near the bottom of the flux tube is because the spline fit departures from the data in these locations. In addition, the absence of any difference in speeds between different passbands (see Fig.~\ref{fig:speeds_simu}), is also a consequence of having a uniform (coronal) temperature in our setup which, again, supports our interpretation of a near isothermal cross-section for the observed loop. Furthermore the values that we calculate from the synthetic maps are comparable to the values found in observations. Therefore, we can attribute the accelerated profile from observations to the projection effect.

\section{Summary and Conclusions}
\label{s:conclusions}

We perform a comprehensive analysis of an oscillatory event observed in an active region fan loop system in the EUV channels of SDO/AIA. We calculate the distribution of periods in the ROI at different heights of the solar atmosphere, from near the photosphere to the corona. We obtain that the oscillations observed in the corona, especially in 171~\AA{}, have a dominant period between 2 and 3 minutes and they are confined to a small region along the fan. By tracking these oscillations down to the transition region, to the chromosphere and near the photosphere, we find that this period is also present and dominant in the sunspot umbra where apparently the footpoints of the coronal fan are rooted. This was previously suggested by \citet{stekel2014}. In addition, the region where the oscillations are localized is expanding towards the corona, indicating an expansion of the waveguide. 
We also perform a kinematic analysis along one of the fan loops observed in the EUV channels 131~\AA{}, 171~\AA{}, 193~\AA{}, and 211~\AA{} of AIA. We calculate the propagation speed of the perturbation in the POS as a function of the distance along the loop, with the origin of the loop in the footpoints located in the umbra of the associated sunspot. We find that the apparent propagation speed is slower than the local sound speed but exhibits an accelerated pattern in all the four channels. Similar acceleration profiles in EUV wavelengths were previously identified and measured by \citet{sheeley2014} and \citet{krishna2017}. But in contrast to \citet{krishna2017}, there is no clear distinction between the speeds determined in the different wavelengths, suggesting a narrow temperature distribution across the loop.

With these observational features as constraints, we perform 2D numerical simulations using the \mancha code. In our numerical model we consider an initial configuration in magnetohydrostatic equilibrium. To achieve this we consider a gravitationally stratified atmosphere using a smooth temperature profile from the photosphere up to the corona. For the magnetic field we use a force-free configuration with a strong magnetic field in the photosphere resembling an  active region. We drive waves in the vertical direction of the momentum equation using a source term with a gaussian shape along the horizontal $x$ and the vertical $z$ axes near the photosphere. The period of the driver is selected to be 3 minutes. Using the forward modelling code FoMo we calculate the intensity in the same four EUV channels, 131\,\AA{}, 171\,\AA{}, 193\,\AA{} and 211\,\AA{}, from SDO/AIA, for a 2D flux tube extracted from within the simulation domain. We then compute the propagation speed of the waves along the magnetic flux tube as seen on the projected $x$-axis, i.e. the apparent speed seen by an observer located above the active region looking along $z$-axis. We obtain accelerated patterns in the wave propagation comparable to the ones detected in the observations with the speed values also in a similar range. 

In observations both thermal and projection effects could be present, producing an accelerated pattern, but in simulations the acceleration is purely due to the projection effect caused by the inclination of the magnetic field. Thus, given the similarities of our numerical and observational results, we attribute the observed accelerated propagation speed to the projection effect. Nevertheless, we emphasize that one should be careful in assigning a physical meaning to the acceleration in the observed propagation speeds of slow waves.

\begin{acknowledgements}
MVS \& TVD acknowledge support from the \emph{European Research Council (ERC)}, grant agreement No 724326 under the European Union's Horizon 2020 research and innovation programme. SKP is grateful to \emph{FWO Vlaanderen} for a senior postdoctoral fellowship (No. 12ZF420N). TVD was also supported by \emph{C1 grant TRACESpace} of Internal Funds KU Leuven. TVD has benefited from the funding of the \emph{FWO Vlaanderen} through a Senior Research Project (G088021N). GS was supported by NASA Grant 80NSSC19K1261. EK acknowledges support from the \emph{Spanish Ministry of Science} through the project PGC2018-095832-B-I00 and from the \emph{European Research Council} through the project ERC-2017-CoG771310-PI2FA. The data used here are courtesy of NASA/SDO and the AIA and HMI science teams. \mancha is a public collaborative code that can be requested from \url{http://research.iac.es/proyecto/PI2FA/pages/codes.php}
\end{acknowledgements}

%
\bibliographystyle{aa} 
\bibliography{biblio} 

\begin{thebibliography}{61}
\expandafter\ifx\csname natexlab\endcsname\relax\def\natexlab#1{#1}\fi

\bibitem[{{Aschwanden}(2004)}]{aschwanden2004}
{Aschwanden}, M.~J. 2004, {Physics of the Solar Corona. An Introduction}
  (Praxis Publishing Ltd., Chichester, UK, and Springer-Verlag Berlin)

\bibitem[{{Banerjee} {et~al.}(2021){Banerjee}, {Krishna Prasad}, {Pant},
  {McLaughlin}, {Antolin}, {Magyar}, {Ofman}, {Tian}, {Van Doorsselaere}, {De
  Moortel}, \& {Wang}}]{banerjee2021}
{Banerjee}, D., {Krishna Prasad}, S., {Pant}, V., {et~al.} 2021, \ssr, 217, 76

\bibitem[{{Berenger}(1994)}]{berenger1994}
{Berenger}, J.-P. 1994, Journal of Computational Physics, 114, 185

\bibitem[{{Berenger}(1996)}]{berenger1996}
{Berenger}, J.-P. 1996, Journal of Computational Physics, 127, 363

\bibitem[{{Berghmans} \& {Clette}(1999)}]{berghmans1999}
{Berghmans}, D. \& {Clette}, F. 1999, \solphys, 186, 207

\bibitem[{{Bogdan}(2000)}]{bogdan2000}
{Bogdan}, T.~J. 2000, \solphys, 192, 373

\bibitem[{{Botha} {et~al.}(2011){Botha}, {Arber}, {Nakariakov}, \&
  {Zhugzhda}}]{botha2011}
{Botha}, G.~J.~J., {Arber}, T.~D., {Nakariakov}, V.~M., \& {Zhugzhda}, Y.~D.
  2011, \apj, 728, 84

\bibitem[{{Chae} {et~al.}(2017){Chae}, {Lee}, {Cho}, {Song}, {Cho}, \&
  {Yurchyshyn}}]{chae2017}
{Chae}, J., {Lee}, J., {Cho}, K., {et~al.} 2017, \apj, 836, 18

\bibitem[{{Cho} {et~al.}(2019){Cho}, {Chae}, {Lim}, \& {Yang}}]{cho2019}
{Cho}, K., {Chae}, J., {Lim}, E.-k., \& {Yang}, H. 2019, \apj, 879, 67

\bibitem[{{De Moortel} {et~al.}(2015){De Moortel}, {Antolin}, \& {Van
  Doorsselaere}}]{demoortel2015}
{De Moortel}, I., {Antolin}, P., \& {Van Doorsselaere}, T. 2015, \solphys, 290,
  399

\bibitem[{{De Moortel} {et~al.}(2002){De Moortel}, {Ireland}, {Hood}, \&
  {Walsh}}]{demoortel2002}
{De Moortel}, I., {Ireland}, J., {Hood}, A.~W., \& {Walsh}, R.~W. 2002, \aap,
  387, L13

\bibitem[{{De Moortel} \& {Nakariakov}(2012)}]{demoortel2012}
{De Moortel}, I. \& {Nakariakov}, V.~M. 2012, Philosophical Transactions of the
  Royal Society of London Series A, 370, 3193

\bibitem[{{De Pontieu} \& {McIntosh}(2010)}]{depontieu2010}
{De Pontieu}, B. \& {McIntosh}, S.~W. 2010, \apj, 722, 1013

\bibitem[{{DeForest} \& {Gurman}(1998)}]{deforest1998}
{DeForest}, C.~E. \& {Gurman}, J.~B. 1998, \apjl, 501, L217

\bibitem[{{Fedun} {et~al.}(2011){Fedun}, {Shelyag}, \&
  {Erd{\'e}lyi}}]{fedun2011}
{Fedun}, V., {Shelyag}, S., \& {Erd{\'e}lyi}, R. 2011, \apj, 727, 17

\bibitem[{{Felipe} {et~al.}(2010){Felipe}, {Khomenko}, \&
  {Collados}}]{felipe2010}
{Felipe}, T., {Khomenko}, E., \& {Collados}, M. 2010, \apj, 719, 357

\bibitem[{{Fleck} \& {Schmitz}(1991)}]{fleck1991}
{Fleck}, B. \& {Schmitz}, F. 1991, \aap, 250, 235

\bibitem[{{Gonz{\'a}lez-Morales} {et~al.}(2019){Gonz{\'a}lez-Morales},
  {Khomenko}, \& {Cally}}]{gonzalez-morales2019}
{Gonz{\'a}lez-Morales}, P.~A., {Khomenko}, E., \& {Cally}, P.~S. 2019, \apj,
  870, 94

\bibitem[{{Gonz{\'a}lez-Morales} {et~al.}(2018){Gonz{\'a}lez-Morales},
  {Khomenko}, {Downes}, \& {de Vicente}}]{gonzalez-morales2018}
{Gonz{\'a}lez-Morales}, P.~A., {Khomenko}, E., {Downes}, T.~P., \& {de
  Vicente}, A. 2018, \aap, 615, A67

\bibitem[{{Hu}(1996)}]{hu1996}
{Hu}, F.~Q. 1996, Journal of Computational Physics, 129, 201

\bibitem[{{Hu}(2001)}]{hu2001}
{Hu}, F.~Q. 2001, Journal of Computational Physics, 173, 455

\bibitem[{{Jess} {et~al.}(2012{\natexlab{a}}){Jess}, {De Moortel},
  {Mathioudakis}, {Christian}, {Reardon}, {Keys}, \& {Keenan}}]{jess2012}
{Jess}, D.~B., {De Moortel}, I., {Mathioudakis}, M., {et~al.}
  2012{\natexlab{a}}, \apj, 757, 160

\bibitem[{{Jess} {et~al.}(2013){Jess}, {Reznikova}, {Van Doorsselaere}, {Keys},
  \& {Mackay}}]{jess2013}
{Jess}, D.~B., {Reznikova}, V.~E., {Van Doorsselaere}, T., {Keys}, P.~H., \&
  {Mackay}, D.~H. 2013, \apj, 779, 168

\bibitem[{{Jess} {et~al.}(2012{\natexlab{b}}){Jess}, {Shelyag}, {Mathioudakis},
  {Keys}, {Christian}, \& {Keenan}}]{jess2012a}
{Jess}, D.~B., {Shelyag}, S., {Mathioudakis}, M., {et~al.} 2012{\natexlab{b}},
  \apj, 746, 183

\bibitem[{{Khomenko} {et~al.}(2009){Khomenko}, {Kosovichev}, {Collados},
  {Parchevsky}, \& {Olshevsky}}]{khomenko2009}
{Khomenko}, E., {Kosovichev}, A., {Collados}, M., {Parchevsky}, K., \&
  {Olshevsky}, V. 2009, \apj, 694, 411

\bibitem[{{Khomenko} {et~al.}(2018){Khomenko}, {Vitas}, {Collados}, \& {de
  Vicente}}]{khomenko2018}
{Khomenko}, E., {Vitas}, N., {Collados}, M., \& {de Vicente}, A. 2018, \aap,
  618, A87

\bibitem[{{Kobanov} {et~al.}(2013){Kobanov}, {Chelpanov}, \&
  {Kolobov}}]{kobanov2013}
{Kobanov}, N.~I., {Chelpanov}, A.~A., \& {Kolobov}, D.~Y. 2013, \aap, 554, A146

\bibitem[{{Krishna Prasad} {et~al.}(2012){Krishna Prasad}, {Banerjee}, \&
  {Singh}}]{krishna2012}
{Krishna Prasad}, S., {Banerjee}, D., \& {Singh}, J. 2012, \solphys, 281, 67

\bibitem[{{Krishna Prasad} {et~al.}(2015){Krishna Prasad}, {Jess}, \&
  {Khomenko}}]{krishna2015}
{Krishna Prasad}, S., {Jess}, D.~B., \& {Khomenko}, E. 2015, \apjl, 812, L15

\bibitem[{{Krishna Prasad} {et~al.}(2017){Krishna Prasad}, {Jess}, {Klimchuk},
  \& {Banerjee}}]{krishna2017}
{Krishna Prasad}, S., {Jess}, D.~B., {Klimchuk}, J.~A., \& {Banerjee}, D. 2017,
  \apj, 834, 103

\bibitem[{{Lemen} {et~al.}(2012){Lemen}, {Title}, {Akin}, {Boerner}, {Chou},
  {Drake}, {Duncan}, {Edwards}, {Friedlaender}, {Heyman}, {Hurlburt}, {Katz},
  {Kushner}, {Levay}, {Lindgren}, {Mathur}, {McFeaters}, {Mitchell}, {Rehse},
  {Schrijver}, {Springer}, {Stern}, {Tarbell}, {Wuelser}, {Wolfson}, {Yanari},
  {Bookbinder}, {Cheimets}, {Caldwell}, {Deluca}, {Gates}, {Golub}, {Park},
  {Podgorski}, {Bush}, {Scherrer}, {Gummin}, {Smith}, {Auker}, {Jerram},
  {Pool}, {Soufli}, {Windt}, {Beardsley}, {Clapp}, {Lang}, \& {Waltham}}]{aia}
{Lemen}, J.~R., {Title}, A.~M., {Akin}, D.~J., {et~al.} 2012, \solphys, 275, 17

\bibitem[{{Marsh} \& {Walsh}(2006)}]{marsh2006}
{Marsh}, M.~S. \& {Walsh}, R.~W. 2006, \apj, 643, 540

\bibitem[{{Marsh} \& {Walsh}(2009)}]{marsh2009b}
{Marsh}, M.~S. \& {Walsh}, R.~W. 2009, \apjl, 706, L76

\bibitem[{{Marsh} {et~al.}(2009){Marsh}, {Walsh}, \& {Plunkett}}]{marsh2009}
{Marsh}, M.~S., {Walsh}, R.~W., \& {Plunkett}, S. 2009, \apj, 697, 1674

\bibitem[{{Misra} {et~al.}(2018){Misra}, {Bora}, \& {Dewangan}}]{misra2018}
{Misra}, R., {Bora}, A., \& {Dewangan}, G. 2018, Astronomy and Computing, 23,
  83

\bibitem[{{Mumford} {et~al.}(2015){Mumford}, {Fedun}, \&
  {Erd{\'e}lyi}}]{mumford2015}
{Mumford}, S.~J., {Fedun}, V., \& {Erd{\'e}lyi}, R. 2015, \apj, 799, 6

\bibitem[{{Nakariakov} \& {Kolotkov}(2020)}]{nakariakov2020}
{Nakariakov}, V.~M. \& {Kolotkov}, D.~Y. 2020, \araa, 58, 441

\bibitem[{{Ofman} \& {Davila}(1997)}]{ofman1997}
{Ofman}, L. \& {Davila}, J.~M. 1997, \apjl, 476, L51

\bibitem[{{Parchevsky} \& {Kosovichev}(2007)}]{parchevsky2007}
{Parchevsky}, K.~V. \& {Kosovichev}, A.~G. 2007, \apj, 666, 547

\bibitem[{{Parchevsky} \& {Kosovichev}(2009)}]{parchevsky2009}
{Parchevsky}, K.~V. \& {Kosovichev}, A.~G. 2009, \apj, 694, 573

\bibitem[{{Pesnell} {et~al.}(2012){Pesnell}, {Thompson}, \& {Chamberlin}}]{sdo}
{Pesnell}, W.~D., {Thompson}, B.~J., \& {Chamberlin}, P.~C. 2012, \solphys,
  275, 3

\bibitem[{Press {et~al.}(2002)Press, Press, Teukolsky, Vetterling, \&
  Flannery}]{press2002numerical}
Press, W., Press, W., Teukolsky, S., Vetterling, W., \& Flannery, B. 2002,
  Numerical Recipes in C++: The Art of Scientific Computing (Cambridge
  University Press)

\bibitem[{{Reznikova} \& {Shibasaki}(2012)}]{reznikovaandshibasaki2012}
{Reznikova}, V.~E. \& {Shibasaki}, K. 2012, \apj, 756, 35

\bibitem[{{Reznikova} {et~al.}(2012){Reznikova}, {Shibasaki}, {Sych}, \&
  {Nakariakov}}]{reznikova2012}
{Reznikova}, V.~E., {Shibasaki}, K., {Sych}, R.~A., \& {Nakariakov}, V.~M.
  2012, \apj, 746, 119

\bibitem[{{Riedl} {et~al.}(2021){Riedl}, {Van Doorsselaere}, {Reale},
  {Goossens}, {Petralia}, \& {Pagano}}]{riedl2021}
{Riedl}, J.~M., {Van Doorsselaere}, T., {Reale}, F., {et~al.} 2021, \apj, 922,
  225

\bibitem[{{Santamaria} {et~al.}(2015){Santamaria}, {Khomenko}, \&
  {Collados}}]{santamaria2015}
{Santamaria}, I.~C., {Khomenko}, E., \& {Collados}, M. 2015, \aap, 577, A70

\bibitem[{{Schou} {et~al.}(2012){Schou}, {Scherrer}, {Bush}, {Wachter},
  {Couvidat}, {Rabello-Soares}, {Bogart}, {Hoeksema}, {Liu}, {Duvall}, {Akin},
  {Allard}, {Miles}, {Rairden}, {Shine}, {Tarbell}, {Title}, {Wolfson},
  {Elmore}, {Norton}, \& {Tomczyk}}]{hmi}
{Schou}, J., {Scherrer}, P.~H., {Bush}, R.~I., {et~al.} 2012, \solphys, 275,
  229

\bibitem[{{Sheeley} {et~al.}(2014){Sheeley}, {Warren}, {Lee}, {Chung}, {Katz},
  \& {Namkung}}]{sheeley2014}
{Sheeley}, N.~R., J., {Warren}, H.~P., {Lee}, J., {et~al.} 2014, \apj, 797, 131

\bibitem[{{Stekel} {et~al.}(2014){Stekel}, {Stenborg}, \& {Dal
  Lago}}]{stekel2014}
{Stekel}, T.~R.~C., {Stenborg}, G., \& {Dal Lago}, A. 2014, in AGU Fall Meeting
  Abstracts, Vol. 2014, SH13B--4110

\bibitem[{{Sych} {et~al.}(2009){Sych}, {Nakariakov}, {Karlicky}, \&
  {Anfinogentov}}]{sych2009}
{Sych}, R., {Nakariakov}, V.~M., {Karlicky}, M., \& {Anfinogentov}, S. 2009,
  \aap, 505, 791

\bibitem[{{Tomczyk} \& {McIntosh}(2009)}]{tomczyk2009}
{Tomczyk}, S. \& {McIntosh}, S.~W. 2009, \apj, 697, 1384

\bibitem[{{Van Doorsselaere} {et~al.}(2016){Van Doorsselaere}, {Antolin},
  {Yuan}, {Reznikova}, \& {Magyar}}]{vandoorsselaere2016}
{Van Doorsselaere}, T., {Antolin}, P., {Yuan}, D., {Reznikova}, V., \&
  {Magyar}, N. 2016, Frontiers in Astronomy and Space Sciences, 3, 4

\bibitem[{{Van Doorsselaere} {et~al.}(2020){Van Doorsselaere}, {Srivastava},
  {Antolin}, {Magyar}, {Vasheghani Farahani}, {Tian}, {Kolotkov}, {Ofman},
  {Guo}, {Arregui}, {De Moortel}, \& {Pascoe}}]{vandoorsselaere2020}
{Van Doorsselaere}, T., {Srivastava}, A.~K., {Antolin}, P., {et~al.} 2020,
  \ssr, 216, 140

\bibitem[{{V{\"o}gler} {et~al.}(2005){V{\"o}gler}, {Shelyag}, {Sch{\"u}ssler},
  {Cattaneo}, {Emonet}, \& {Linde}}]{vogler2005}
{V{\"o}gler}, A., {Shelyag}, S., {Sch{\"u}ssler}, M., {et~al.} 2005, \aap, 429,
  335

\bibitem[{{Wang}(2016)}]{wang2016}
{Wang}, T.~J. 2016, Washington DC American Geophysical Union Geophysical
  Monograph Series, 216, 395

\bibitem[{{Wang} {et~al.}(2009){Wang}, {Ofman}, {Davila}, \&
  {Mariska}}]{wang2009}
{Wang}, T.~J., {Ofman}, L., {Davila}, J.~M., \& {Mariska}, J.~T. 2009, \aap,
  503, L25

\bibitem[{{Yuan} \& {Nakariakov}(2012)}]{yuan2012}
{Yuan}, D. \& {Nakariakov}, V.~M. 2012, \aap, 543, A9

\bibitem[{{Yuan} {et~al.}(2014){Yuan}, {Sych}, {Reznikova}, \&
  {Nakariakov}}]{yuan2014}
{Yuan}, D., {Sych}, R., {Reznikova}, V.~E., \& {Nakariakov}, V.~M. 2014, \aap,
  561, A19

\bibitem[{{Zhao} {et~al.}(2016){Zhao}, {Felipe}, {Chen}, \&
  {Khomenko}}]{zhao2016}
{Zhao}, J., {Felipe}, T., {Chen}, R., \& {Khomenko}, E. 2016, \apjl, 830, L17

\bibitem[{{Zhukov}(2002)}]{zhukov2002}
{Zhukov}, V.~I. 2002, \aap, 386, 653

\bibitem[{{Zurbriggen} {et~al.}(2020){Zurbriggen}, {Sieyra}, {Costa},
  {Esquivel}, \& {Stenborg}}]{zurbriggen2020}
{Zurbriggen}, E., {Sieyra}, M.~V., {Costa}, A., {Esquivel}, A., \& {Stenborg},
  G. 2020, \mnras, 494, 5270

\end{thebibliography}
%
\end{document}